\documentclass[
  journal=pasa,
  manuscript=research-paper, 
  year=202X,
  volume=YY,
]{cup-journal}

\usepackage{graphicx}
\usepackage[export]{adjustbox}
\usepackage{lscape}
\usepackage{color}
\usepackage{booktabs}
\usepackage{amsmath}
\usepackage{caption}
\usepackage{subcaption}
\usepackage{aas-macros}
\usepackage{hyperref}
\usepackage{orcidlink}
\usepackage{pdflscape}
\hypersetup{colorlinks,citecolor=blue,linkcolor=blue,urlcolor=blue}
\usepackage{amssymb,microtype,siunitx,booktabs,multirow}
\usepackage[normalem]{ulem}

\usepackage{placeins}
\usepackage{array} 

\newcommand\kms{km\,s$^{-1}$}

\newcommand\hi{\mbox{H\,{\sc i}}\ }

\title{ASKAP-FLASH: An \hi absorption search in compact radio sources from the MWA IPS survey}

\author{Samhita S. Sodhi}
\affiliation{Sydney Institute for Astronomy, School of Physics A28, University of Sydney, NSW 2006, Australia}
\email[Samhita S. Sodhi]{ssod4755@uni.sydney.edu.au}

\author{Elaine M. Sadler}
\affiliation{Sydney Institute for Astronomy, School of Physics A28, University of Sydney, NSW 2006, Australia}
\alsoaffiliation{ATNF, CSIRO, Space and Astronomy, PO Box 76, Epping, NSW 1710, Australia}

\author{Rajan Chhetri}
\affiliation{ATNF, CSIRO, Space and Astronomy, PO Box 1130, Bentley WA 6102, Australia} 

\author{Emily F. Kerrison}
\affiliation{Sydney Institute for Astronomy, School of Physics A28, University of Sydney, NSW 2006, Australia}
\alsoaffiliation{ATNF, CSIRO, Space and Astronomy, PO Box 76, Epping, NSW 1710, Australia}

\author{Elizabeth K. Mahony}
\affiliation{ATNF, CSIRO, Space and Astronomy, PO Box 76, Epping, NSW 1710, Australia}

\author{John Morgan}
\affiliation{ATNF, CSIRO, Space and Astronomy, PO Box 1130, Bentley WA 6102, Australia} 

\author{Angie Waszewski}
\affiliation{School of Science, University of Newcastle, Callaghan, NSW 2308, Australia} 

\author{Hyein Yoon}
\affiliation{Korea Astronomy and Space Science Institute, 776 Daedeokdae-ro, Yuseong-gu, Daejeon 34055, Republic of Korea}
\alsoaffiliation{Sydney Institute for Astronomy, School of Physics A28, University of Sydney, NSW 2006, Australia}

\author{J.N.H.S. Aditya}
\affiliation{Shanghai Astronomical Observatory, Chinese Academy of Sciences, China}

\author{James R. Allison}
\affiliation{First Light Fusion Ltd., Unit 9/10 Oxford Pioneer Park, Yarnton, OX5 1QU, UK}

\author{Ron Ekers}
\affiliation{ATNF, CSIRO, Space and Astronomy, PO Box 76, Epping, NSW 1710, Australia}

\author{Marcin Glowacki}
\affiliation{
Institute for Astronomy, University of Edinburgh, Royal Observatory, Edinburgh, UK} 
\alsoaffiliation{Department of Astronomy, Inter-University Institute for Data Intensive Astronomy, University of Cape Town, Cape Town, South Africa} 

\author{Paris E. Gordon-Hall}
\affiliation{Sydney Institute for Astronomy, School of Physics A28, University of Sydney, NSW 2006, Australia}
\alsoaffiliation{ATNF, CSIRO, Space and Astronomy, PO Box 76, Epping, NSW 1710, Australia}

\author{Vanessa Moss}
\affiliation{ATNF, CSIRO, Space and Astronomy, PO Box 76, Epping, NSW 1710, Australia}

\author{William Roster}
\affiliation{Max Planck Institute for Extraterrestrial Physics, 85748 Garching near Munich, Germany}

\author{Mara Salvato}
\affiliation{Max Planck Institute for Extraterrestrial Physics, 85748 Garching near Munich, Germany}

\author{Stanislav S. Shabala }
\affiliation{School of Natural Sciences, University of Tasmania, Private Bag 37, Hobart, TAS, 7001, Australia}

\author{Renzhi Su}
\affiliation{Shanghai Astronomical Observatory, Chinese Academy of Sciences, China}

\author{Ross J. Turner} 
\affiliation{School of Natural Sciences, University of Tasmania, Private Bag 37, Hobart, TAS, 7001, Australia}

\author{Matthew Whiting}
\affiliation{ATNF, CSIRO, Space and Astronomy, PO Box 76, Epping, NSW 1710, Australia}

\received {dd Mmm YYYY}
\revised  {dd Mmm YYYY}
\accepted {dd Mmm YYYY}
\published{dd Mmm 202X}

\begin{document}

\begin{abstract}
The wide-field of view and radio-quiet site of the ASKAP radio telescope has enabled large-area, untargeted searches for 21\,cm \hi absorption against distant radio sources. The First Large Absorption Survey in \hi (FLASH) probes the intermediate redshift range between $z=0.42$ and $z=1.0$ where \hi absorption searches have previously been sparse. The aim of this study was to test whether associated \hi absorption is more common in peaked-spectrum radio sources than in compact radio sources with flat or power-law SEDs. This might be expected, for example, if the peaked radio spectrum arises from young radio jets drilling through a dense central cocoon of gas. 
We used interplanetary scintillation (IPS) measurements at 162\,MHz from the Murchison Widefield Array (MWA) to create a catalogue of 711 low-frequency compact radio sources with at least 80\% of their total flux density contained within a region smaller than $\sim0.3$ arcsec in angular size. Two-thirds of these compact sources (61\%) have peaked-spectrum radio SEDs and one-third (32\%) have a flat or power-law radio spectrum. 
We then used FLASH spectral-line data to search the 268 brightest sources (856\,MHz flux density $>30$\,mJy) for redshifted \hi absorption lines at $0.42<z<1.0$. Eight \hi absorption lines were detected, with peak optical depth $\tau_{\rm pk}$ ranging from 0.015 to 0.277. All eight detections were in sources with a peaked radio spectrum. However, we also found that peaked radio spectra were almost ubiquitous among the brightest compact sources in our sample (in the redshift range  $0.42<z<1.0$, over 90\% of IPS selected compact sources brighter than 300\,mJy have peaked-spectrum radio SEDs).
As a result, we were not able to make a robust statistical comparison between the \hi properties of compact radio sources with and without a spectral peak. We find an \hi detection rate of 16$\pm$8\% for associated absorption lines with peak optical depth $\tau_{\rm pk}\geq0.03$, and our results are consistent with a picture in which many peaked-spectrum radio sources contain a central reservoir of cold and dense neutral gas. 
\end{abstract}

\section{Introduction}
\label{sec:int}
21\,cm \hi absorption is a useful tracer of neutral gas in distant galaxies since (unlike the case for \hi emission) the detection rate is independent of redshift and is set mainly by the brightness of the background radio continuum source used as a probe. 

Associated \hi absorption, which arises from gas located within the host galaxy of the radio source, can trace the AGN fuelling and feedback processes that shape the evolution of massive galaxies across cosmic time \citep[e.g.][]{croton2006,morganti2015}. It has been known for many years \citep{vermeulen2003,Morganti2026} that the observed detection rate of associated \hi absorption is higher in compact radio sources (less than a few kpc in size) than in more extended sources. Targeted \hi absorption surveys of nearby compact radio galaxies have achieved detection rates for associated \hi absorption as high as 25--30\% \citep{Allison2012,gereb2015,maccagni2017}, implying that neutral gas is commonly present in the inner regions of galaxies that host compact radio sources. Intervening \hi absorption, on the other hand, occurs along the line of sight towards the background radio continuum source which provides information about the neutral gas distribution in the foreground of the radio source \citep{Brown1973,Su2022}. The detectability for intervening \hi absorption lies between 5--15\% for quasar-galaxy systems in the local Universe and up to $\sim$50\% in redshifted Damped Lyman-$\alpha$ systems \citep{Allison2022}.

Extended radio sources typically show broken or power-law radio spectra, where  
the flux density decreases at higher frequencies due to a smaller number of highly energetic radiating electrons \citep{conway1963,kellermann1964}. 
In contrast, many compact radio sources display flat or peaked radio spectra \citep{O'Dea2021}. 
There are two theories for the spectral turnover in peaked-spectrum sources: synchrotron self-absorption (SSA) or free–free absorption (FFA), both of which can produce a turnover at low radio frequencies \citep{Kellermann1966,ODea1998,Callingham2017}. Some compact sources may also exhibit steep or flat-spectra whose turnover frequencies lie outside the observed radio band \citep{Chhetri2018,Kerrison2024}. 

Peaked-spectrum radio sources are widely interpreted as young radio galaxies, in the early stages of jet expansion that may subsequently evolve into large-scale systems \citep{Philstrom2003,Chandola2011,Orienti2016}. The peaked-spectrum phase is likely to be short-lived, lasting between about 1000\,years and 1 Myr \citep{Kukreti2024}. The compact sizes of these sources in the alternative `frustrated' scenario may reflect interactions with dense surrounding gas causing restriction and confinement \citep{Morganti2018,O'Dea2021}. As the jets expand beyond the host galaxy, the radio spectrum generally evolves toward a power-law form lasting around tens of Myr \citep{Bicknell2018,Ballieux2024,Kukreti2024,Young2025}.

Detections of associated \hi absorption are known to be more common in compact sources than in extended sources \citep{Gupta2006}. This is generally attributed to a higher \hi covering factor in compact sources \citep{Morganti2018,O'Dea2021}.  
Retriggered radio sources are seen when the AGN has undergone multiple episodes of nuclear activity, producing both an older steep-spectrum power-law component and a younger peaked-spectrum component. In these sources, the resulting radio spectral energy distribution (SED) is a combination of these two components \citep{Callingham2017,Kerrison2024}.

The ASKAP First Large Absorption Survey in \hi \citep[FLASH;][]{Allison2022} is an untargeted search for \hi absorption (at redshifts $0.42< z < 1.0$) against bright (flux density $>30$\,mJy) radio sources across 24,000 deg$^2$ of sky, centred at 856\,MHz. The FLASH Pilot Survey \citep{yoon2025} detected 30 new \hi absorption lines in a sky area of 3000 deg$^2$. 67\% of these sources had peaked radio spectra, a clear excess over the general radio-source population at low-frequencies, where only $\sim$5--6\% of sources have a peaked radio spectrum \citep[][Chhetri et al. in prep]{Callingham2017}. 

An interesting question, therefore, is whether peaked-spectrum sources are common in the \cite{yoon2025} sample just because they are compact sources (which are known to have a high \hi detection rate), or whether they are over-represented because they are more \hi-rich than other compact radio sources with flat or power-law spectra. To examine this, we select a sample of compact radio sources through the use of interplanetary scintillation (IPS), and then study both the radio SEDs and the \hi absorption properties of these compact sources.  

Section \ref{sec:sel} describes the selection of the compact source sample using IPS measurements from the MWA radio telescope \citep{Morgan2018,Chhetri2018} in a region of sky that was also covered by early observations in the ASKAP FLASH survey
\citep{Allison2022,yoon2025}. 

In Section \ref{sec:radioSED}, we use the Bayesian RadioSED modelling framework developed by \cite{Kerrison2024} to classify the radio SEDs as peaked or non-peaked. This is followed by Section \ref{sec:props} which matches the sources with FLASH optical counterparts from the sample in \cite{Roster2026}. 

Finally, in Section \ref{sec:flash} we search for \hi absorption in the redshift range $0.42<z<1.0$ and conduct an analysis of the detections found in Section \ref{sec:analysisabsorbers}. Section \ref{sec:discussion} discusses the findings and implications of the study, and the conclusion is presented in Section \ref{sec:conclusion}. 

\section{Selection of the compact source sample}
\label{sec:sel}
\subsection{Interplanetary Scintillation (IPS)}
The region surrounding the Sun consists of an ionised, inhomogeneous solar wind plasma that causes passing electromagnetic waves to be scattered and refracted. This phenomenon, known as interplanetary scintillation (IPS), produces brightness fluctuations in radio sources with sub-arcsecond structure on timescales of a few seconds \citep{clarke1964,Hewish1964,Morgan2018}. IPS therefore provides a powerful method for identifying compact radio sources simultaneously over large areas of sky compared to other commonly used methods such as Very Long Baseline Interferometry (VLBI) \citep{Morgan2018,Chhetri2018}.

The scintillation index, $m$, is defined as the ratio of the scintillating flux density to the mean flux density of the source. Because scintillation depends on solar elongation and scattering conditions, $m$ must be normalised to account for these effects \citep{Chhetri2018,Morgan2019}. 

The Normalised Scintillation Index (NSI) values are expected to lie between 0 and 1, though sources with NSI $>$ 1 can sometimes arise due to solar wind variation, or for sources that are intrinsically variable on timescales of seconds, such as pulsars \citep{Rickett2006,Chhetri2018}. 

An NSI value close to 1 indicates that the source is a point source, and for the MWA at 162\,MHz, this implies an angular diameter smaller than $\sim$0.3 arcsec \citep{waszewski2026}. In contrast, an NSI value close to 0 indicates an extended source (compared to the beam size) where little or no flux arises from a compact component. 

\subsection{Using IPS to select compact radio sources}
To select a sample of compact radio sources that could be searched for \hi absorption, we combined the most recent MWA IPS catalogue at 162\,MHz \citep{Morgan2022, waszewski2026} with a catalogue of radio continuum sources from 71 fields observed in the FLASH survey between November 2022 and October 2024 and downloaded from the CSIRO ASKAP Data Archive (CASDA). 

The final IPS survey catalogue \citep{Morgan2022, waszewski2026} is based on multiple observations of each source in order to average over variations in the solar wind.

\begin{figure*}[t]
    \centering
    \includegraphics[width=0.95\linewidth]{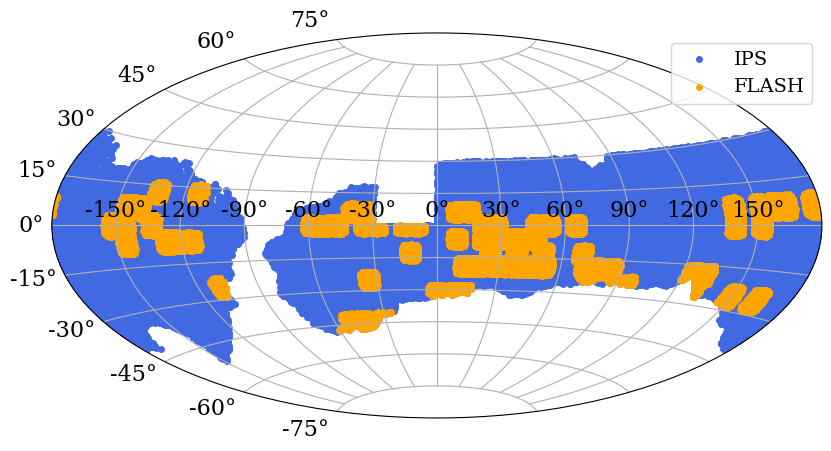}
    \caption{The sky coverage of the FLASH (orange) and IPS (blue) datasets used to create the compact sample. The coordinates are right ascension (RA) versus declination (DEC). FLASH observations which did not overlap the IPS dataset were excluded from the study, and the overlapping data points which introduces duplicate sources, were removed.} 
    \label{fig:overlapflaships}
\end{figure*}

The FLASH and IPS catalogues were restricted to a common sky area (Figure \ref{fig:overlapflaships}) and crossmatched to associate an NSI value with each FLASH radio source.

Because small positional offsets arise from statistical and instrumental uncertainties, crossmatching requires the adoption of a separation threshold. This approach is commonly used in radio catalogue matching \citep{Prescott2016,Ching2017,Kerrison2024}. 

\begin{figure}
    \centering
    \includegraphics[width=0.98\linewidth]{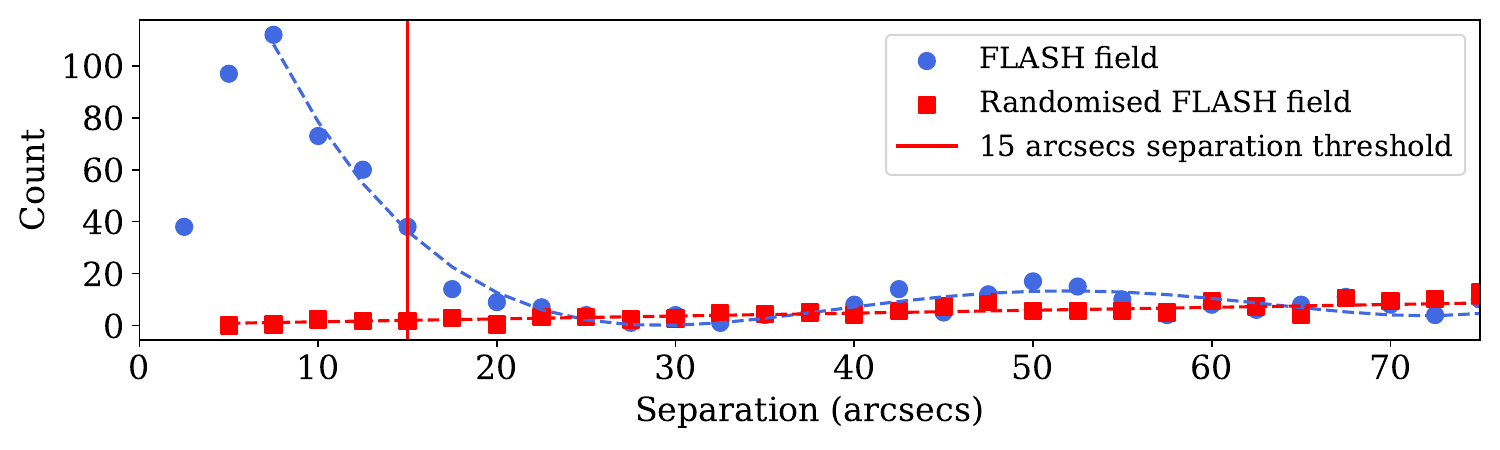}
    \caption{The number of total matches between the FLASH and IPS datasets (y-axis) against the separation between the radio sources (x-axis) for a single FLASH field (SBID 51441). The red vertical line corresponds to the 15 arcsecond threshold. The crossmatching between the original FLASH field and the IPS dataset is in blue filled circles, and randomised FLASH field with varied DEC values for the radio sources in red squares, both with a line of best fit.}
    \label{fig:crossmatching}
\end{figure}

To determine an appropriate threshold, we selected and created a randomised FLASH field by artificially shifting the declination value by 1--2 degrees. We then used TopCat \citep{Taylor2005} to crossmatch both the real FLASH field and the randomised field with the IPS dataset, using an arbitrary large threshold of 75 arcsec. Figure \ref{fig:crossmatching} shows the result, and the convergence of the two curves (at a separation around 25\,arcsec) marks the point at which real and false matches become equally likely. 

After repeating this procedure for several FLASH fields, a threshold of 15 arcsec was chosen. This value retains a relatively large sample of sources while maintaining a high level of reliability for the sources selected, since a smaller separation threshold increases reliability at the expense of completeness. 

Applying this threshold yielded an initial sample of 21,224 matched sources. To define the compact sample, we selected only sources with NSI $\geq$ 0.8, i.e. sources that scintillate at $\geq$80\% of the level expected for a point source less than 0.5\, arcsec in angular size at 162\,MHz (i.e. a projected linear size $\lesssim$ 2--3\,kpc in the FLASH redshift range). This cut-off is slightly less restrictive than the value of NSI $>$ 0.9 adopted by \cite{Chhetri2018}. 
Our NSI $\geq$ 0.8 criterion yielded 818 compact sources, compared to only 369 sources for a cut-off of NSI $\geq$ 0.9. We then removed duplicate sources detected in the overlapping regions of adjacent FLASH fields, giving us a final sample of 711 IPS compact sources, containing 71 FLASH fields each covering a field of view of 6.4 $\times$ 6.4 degrees.

\section{Radio SED fitting}
\label{sec:radioSED}
\begin{figure*}
     \centering
             \subfloat[Peaked Spectrum]{\includegraphics[width=0.32\textwidth]{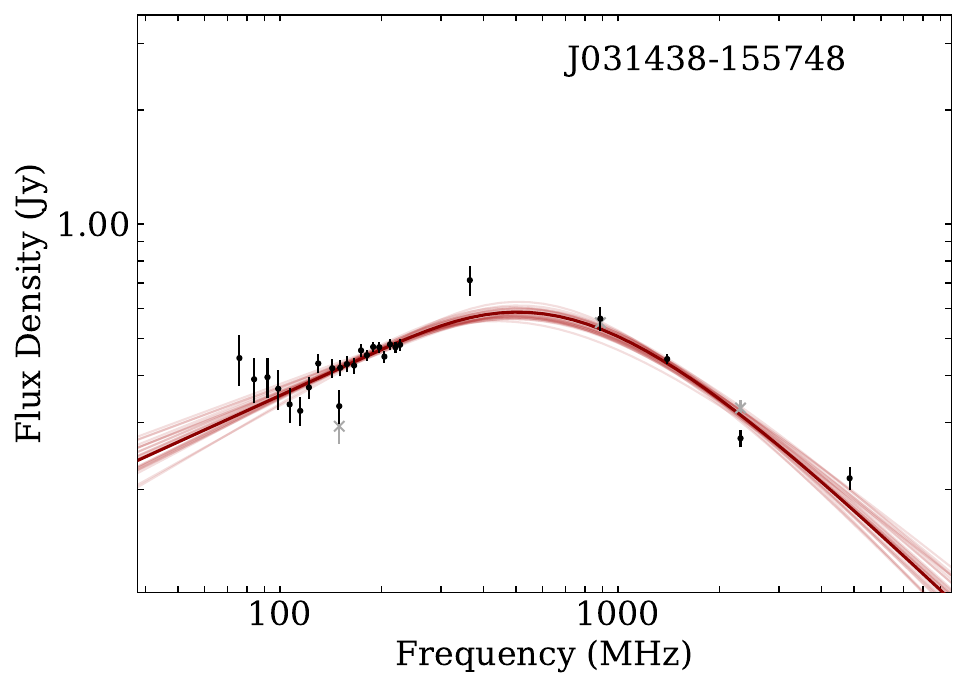}}
                     \subfloat[Power-Law Spectrum]{\includegraphics[width=0.32\textwidth]{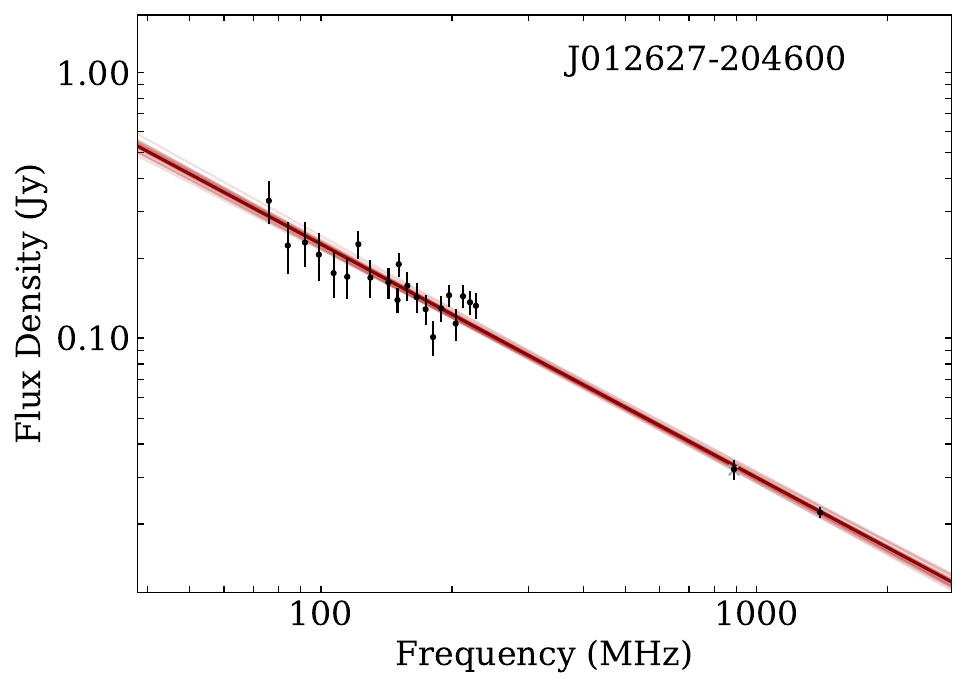}}
     \subfloat[Retriggered Spectrum]{\includegraphics[width=0.32\textwidth]{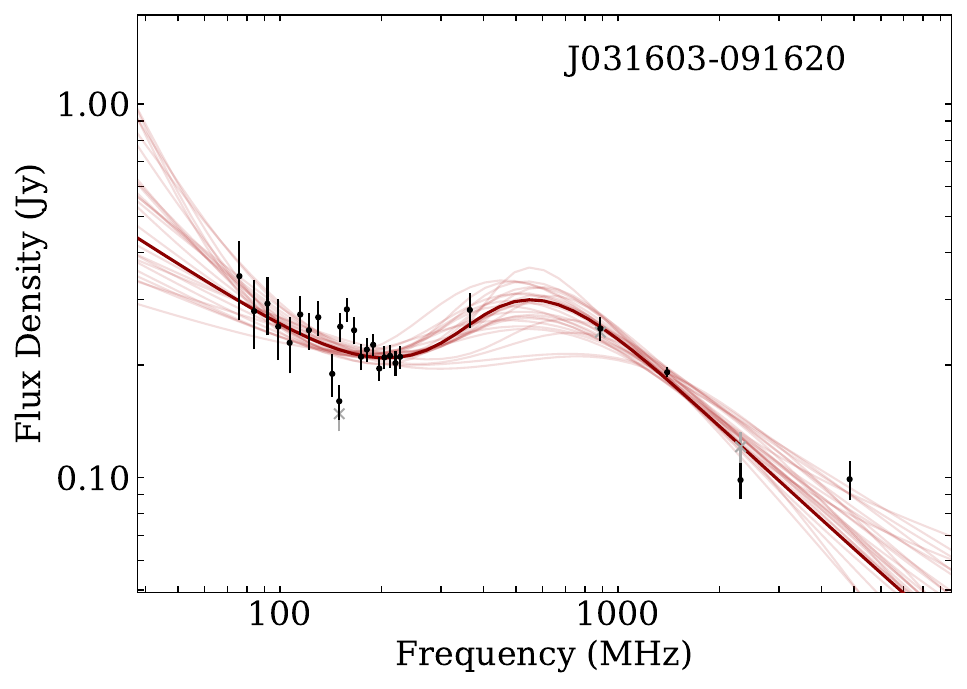}}
     \caption{Examples of the three spectral classes for the radio SED fits. (a) Source J031438-155748 is fitted by a peaked spectrum. (b): Source J012627-204600 is fitted as a power-law spectrum. (c): Source J031603-091620 follows a retriggered spectrum. The faint red lines reveal the samples taken from the posterior distribution by the Bayesian RadioSED package.}
     \label{fig:combinedsed}
\end{figure*}

The radio SED fitting was performed using a Bayesian inference package RadioSED \footnote{\href{https://github.com/ekerrison/RadioSED/}{https://github.com/ekerrison/RadioSED/}} \citep{Kerrison2024}, which uses the large-area Rapid ASKAP Continuum Survey (RACS)-low \citep{hale2021} as a reference catalogue. Using the classification methods of  \cite{Kerrison2024}, we tested three models for the radio spectral energy distribution of an individual radio source --- peaked spectrum, power-law spectrum (including flat-spectrum sources) and retriggered spectrum, see Figure \ref{fig:combinedsed}. The package uses non-informative priors as appropriate for each parameter; a uniform prior for spectral indices and a Jeffreys' prior for peak frequency and flux density (since fitting is performed in log-space). The priors for peak frequency and flux density are constrained within the parameter space spanned by the surveys incorporated into fitting (currently $50\,\rm{MHz} <$ $\nu_p < 100\,\rm{GHz}$ and $30\,\rm{mJy} < S_p < 15\,\rm{Jy}$). The Bayes Factor derived for each fit is a measure of how well one model performs compared to another, and indicates the best fitting model. 

SED fits were derived for 706 of the 711 compact sources.  Five sources were not fitted, either because there was not enough data or because the source was not found in the RACS-low catalogue. Some of the fits were found to have a Bayes Factor between 0 and 0.5 which suggested the preferences between the models were marginal due to fits affected by data from older surveys and also intrinsic variability of the sources. Thus, some of the spectral classes were recategorised. The results did not have a significant difference but an extra category labelled `Uncertain' was added for sources where the best fitting model could not be determined. These results are summarised in Table \ref{tab:compactsedresults}. 

In this sample of compact sources, peaked-spectrum SEDs are almost twice as common as compact power-law spectrum sources, 61\% compared to 32\%. Only 2\% of sources had a retriggered spectrum. Our results agree with those of \cite{Chhetri2018}, who found that peaked-spectrum radio sources were the dominant population among the strongly scintillating, low-frequency sources. 

In contrast, \cite{Ballieux2024} estimated that peaked-spectrum sources make up less than 5\% of the general population of low-frequency radio sources. A similar result was found by \cite{paris2026} for a sample of 99 extended low-frequency sources with IPS values of NSI $<$ 0.8. Around 5\% of these sources had peaked spectra, and none of the sources with an NSI value below 0.4 showed a spectral peak. 

In our compact sample, therefore, peaked-spectrum sources are over ten times more common than in the general radio-source population. 

For the \hi absorption search, we further reduced the compact source sample by limiting it to sources with a peak flux density above 30\,mJy\,beam$^{-1}$, since \hi absorption is unlikely to be detectable by FLASH for fainter sources \citep{yoon2025}. In this reduced sample of 268 sources (Table \ref{tab:compactsedresults}), peaked-spectrum sources are still the dominant population. 

\begin{table}
    \centering
\caption{Results of the radio SED modelling performed on the full sample of 706 compact sources with NSI $\geq$ 0.8 in the middle column and the sample of 268 bright compact sources ($>$30 mJy\,beam$^{-1}$) in the right-hand column.}
\label{tab:compactsedresults}
    \begin{tabular}{l>{\raggedleft\arraybackslash}p{2.5cm}>{\raggedleft\arraybackslash}p{2.5cm}}
    \toprule
         \textbf{Spectral Classes}& \textbf{Full sample of}&\textbf{Reduced sample of}\\
 & \textbf{706 sources}&\textbf{268 sources}\\
         \hline
          \textbf{Peaked Spectrum}& 432 (61.1\%)& 193 (72.0\%)\\
         \textbf{Power-law Spectrum}& 229 (32.4\%)&55 (20.5\%)\\
         \textbf{Retriggered Spectrum}& 15 (2.1\%)&10 (3.7\%)\\
         \textbf{Uncertain}& 30 (4.2\%)&10 (3.7\%)\\
 \textbf{Total sources}&706&268\\ \bottomrule
    \end{tabular}
\end{table}
 
\section{Optical counterparts of the compact source sample}
\label{sec:props} 
Next, we matched the sample of 268 bright, compact sources with the catalogue of FLASH optical counterparts published by \citet{Roster2026}. This work applies machine learning methods to images from the DESI Legacy Imaging Surveys DR10 \citep[LS10;][]{Dey2019} to identify the most likely optical counterpart for each radio source. 
\cite{Roster2026} also estimated a photometric redshift for each optical counterpart using the PICZL code \citep{Roster2024} and a training set optimised for active galaxies. 

\cite{Roster2026} only made optical identifications for single-component FLASH sources, but 26 of the 268 sources have two components listed in the FLASH continuum catalogue, and so were not included in the \cite{Roster2026} catalogue. 
For these objects, we list the \cite{zhou2021} DR9 photometric redshift for the closest optical counterpart within 5\,arcsec of the radio position. 

For 29 sources where good-quality photometry was available, no optical counterpart could be seen within 5\,arcsec of the radio position. These optical `blank fields' were assumed to correspond to distant galaxies at redshift $z>1.0$.  

Eleven objects are marked as `Uncertain' in Table \ref{tab:opt}, and of these are in fields without LS10 optical photometry. There are two objects where LS10 images are available but the optical identification was uncertain. In one case (NVSS\,J020704-034909) there is a bright foreground star close to the radio position, and the other object (NVSS\,J231740-125606) has two possible counterparts equally close to the radio position. 

Spectroscopic redshifts were available for 47 of the 268 sources in the compact sample, either from the DESI DR1 spectroscopic survey \citep{DESI2026} or from published results. These redshifts are listed in Table \ref{tab:WRcrossmatch} along with the source of each redshift measurement. 
 
We used these spectroscopic redshifts, along with the photometric redshift estimates, to place the sources into the three ranges shown in Table \ref{tab:opt}. 
From this table, we can see that over half of the bright compact sources in our sample are at redshift $z>1.0$. These are background sources for the FLASH redshift range, where we could detect intervening \hi absorption lines but not associated lines. 94 sources lie at $0.42<z<1.0$, the redshift range in which detection of both associated lines (the main focus of this paper) and intervening lines is possible. For the 18 sources with $z<0.42$, i.e. below the FLASH redshift range, detections of an \hi line will not be possible.  

\begin{table}[t]
    \centering
\caption{Redshift distribution of the 268 objects in the bright, compact source sample.}
\label{tab:opt}
    \begin{tabular}{llrrrr}
    \toprule
         \multicolumn{2}{c}{Redshift range} & Galaxy & Quasar & Faint/ &  Total \\
         & & & & blank & \\
         \hline
          Foreground & $z<0.42$ & 18 & .. & .. & 18 \\ 
         In-band & $0.42<z<1.0$ & 84 &  10 & .. & 94 \\
         Background & $z>1.0$ & 68 & 48 & 29 & 145 \\
         Uncertain & &  & & & 11 \\
         Total & & 170 & 58 & 29 &  \multicolumn{1}{r}{268}\\ \bottomrule
    \end{tabular}
\end{table}

\section{\hi absorption search}
\label{sec:flash}
\subsection{Search method}
To search for \hi absorption, we applied the FLASHfinder{\footnote{\href{https://github.com/drjamesallison/flash_finder}{https://github.com/drjamesallison/flash\_finder}}} automated Bayesian line-finding tool \citep{Allison2012} to the FLASH spectra for the 268 compact sources with flux density brighter than 30\,mJy. 

This tool searches for potential \hi lines using a Monte Carlo multi-nested sampling algorithm to determine the best fitting parameters for a Gaussian curve fitted to potential absorption lines in the spectra, and assigns a statistical significance to the detections. No redshift priors were used in the line search.

For detected lines, FLASHfinder outputs the redshift of the peak of the line $z_{\rm HI}$, peak optical depth $\tau_{\rm pk}$, integrated optical depth $\tau_{\rm int}$ in \kms, line width in \kms and a Bayesian evidence value ln\,(B) which provides a numerical confidence value on the line. To make our final line list, we used the same cut-off value of ln\,(B) $>$ 30 as \cite{yoon2025}. 
The detection limit for \hi absorption varies with the brightness of the continuum source used as a probe, and is listed for each individual source in Table \ref{tab:WRcrossmatch}. For a 185\,mJy source (the median value for the objects in Table \ref{tab:WRcrossmatch}), the upper limit in peak optical depth is $\tau_{\rm pk}>$ 0.09, and for a 500\,mJy source the limit is $\tau_{\rm pk}>$ 0.03.

Table \ref{tab:WRcrossmatch} lists the sample of 268 compact sources that were searched for \hi absorption using the FLASH data. Only the first 20 lines are shown, with the full table appearing as an electronic table in the online version of the paper. 

\subsection{\hi line detections} 
We detected \hi absorption against eight of the 268 sources listed in Table \ref{tab:WRcrossmatch}. 
All eight of the \hi detections were against bright radio sources (with 856\,MHz continuum flux densities above 500\,mJy), and all eight detected sources have peaked radio spectra. There were no \hi detections from the other SED classes. 

Table \ref{tab:HIdata} lists the measured properties of the detected lines, and the \hi absorption profiles are shown in \ref{sec:lines}. 

The median value of the \hi peak optical depth $\tau_{\rm pk}$ for sources in Table \ref{tab:HIdata} is 0.052, which is intermediate between the median value of $\tau_{\rm pk}=0.148$ found by \cite{yoon2025} for FLASH Pilot Survey sources and the median $\tau_{\rm pk}=0.030$ found by \cite{Curran2021} for a sample of lines compiled from published results. \cite{yoon2025} note that FLASH, as a relatively shallow survey, is expected to detect lines with higher optical depth than those found in existing studies of individual sources. The different values for our sample and that of \cite{yoon2025} arise from the additional NSI constraints used to select our compact sources, which skew our sample towards brighter sources as we discuss in \S\ref{sec:discussion}. 

The median \hi line width for the sources in Table \ref{tab:HIdata} is 67\,\kms, similar to the median value of 79\,\kms\ for the \cite{yoon2025} Pilot Survey sample. 

The overall \hi detection rate for the compact sample is 8$/$268, or 3.0 $\pm$ 0.9\%, but this does not take into account the redshift distribution of our sources or the fact that absorption lines are harder to detect against fainter radio sources \citep[see e.g. Fig. 19 of ][]{yoon2025}. We discuss these selection effects further in section \S\ref{sec:selection effects} and derive a final \hi detection rate for associated absorption lines in \S\ref{sec:detrate}.

\subsection{Associated or intervening \hi lines?}
We can now look at whether the detected lines arise from associated or intervening absorption (see Column 18 in Table \ref{tab:HIdata}).

In six of the eight detections, we classify the \hi lines as associated absorption systems. The evidence for this is both the close match between the optical and \hi redshifts and the relatively broad widths of the majority of these absorption lines. 

As noted by \cite{Curran2021}, associated \hi absorption lines typically have much broader line widths than intervening lines. The median width of 91.5\ \kms\ for the six associated lines in Table \ref{tab:HIdata} is close to the median value of 84.5\,\kms\ for associated lines in the \cite{Curran2021} sample. 

We classify the two narrowest lines in Table \ref{tab:HIdata}, detected against PKS 2127+04 and MRC\,2131-241, as intervening systems. These two objects are different in nature, and we discuss each of them separately below. 

\subsubsection{MRC\,2131-241}
MRC\,2131-241 is a bright radio source in the MRC-1Jy sample \citep{mccarthy1996}, but lacks an optical spectroscopic redshift. \cite{Roster2026} estimate a photometric redshift of z = 0.66, though with a large uncertainty (range 0.18 to 1.86) because only two LS10 photometric bands are available. 

This object is discussed in detail by \cite{paris2026}, who used interplanetary scintillation data at 820\,MHz from pilot IPS observations with the ASKAP telescope \citep{Chhetri2026arXiv}. Their ASKAP NSI value of 0.84 $\pm$ 0.04 is very close to the MWA value of 0.85 $\pm$ 0.05, implying that the compact component is smaller than 0.1\,arcsec in angular size. 

In the absence of a spectroscopic redshift, \cite{paris2026} argue that this is likely to be an intervening \hi line, and that the radio counterpart is a faint background object originally identified by 
\cite{mccarthy1996} and offset from the brighter galaxy seen in LS10 images. An optical spectrum would be valuable to confirm this. 

\subsubsection{PKS\,2127+04}
For PKS\,2127+04, the \hi redshift of 0.9088 $\pm$ 0.0001 is significantly lower than the optical spectroscopic redshift of 0.99 published by \cite{stickel1994}. No details of the optical spectrum are available, but \cite{stickel1994} note that ``The extremely faint optical counterpart has been found by \cite{biretta1985}. The redshift was made available to us by H. Spinrad (1993, priv. comm)''. On this basis, we classify PKS\,2127+04 as an intervening \hi system. 

PKS\,2127+04 is the brightest radio source in our sample, and has been extensively studied at radio wavelengths. New optical data, both imaging and spectroscopy, would be valuable to search for the intervening galaxy. 

\section{Properties of the absorbers}
\label{sec:analysisabsorbers}

\subsection{Linear sizes of the compact sources}
\label{sec:size}
\begin{figure}
    \centering
    \includegraphics[width=0.98\linewidth]{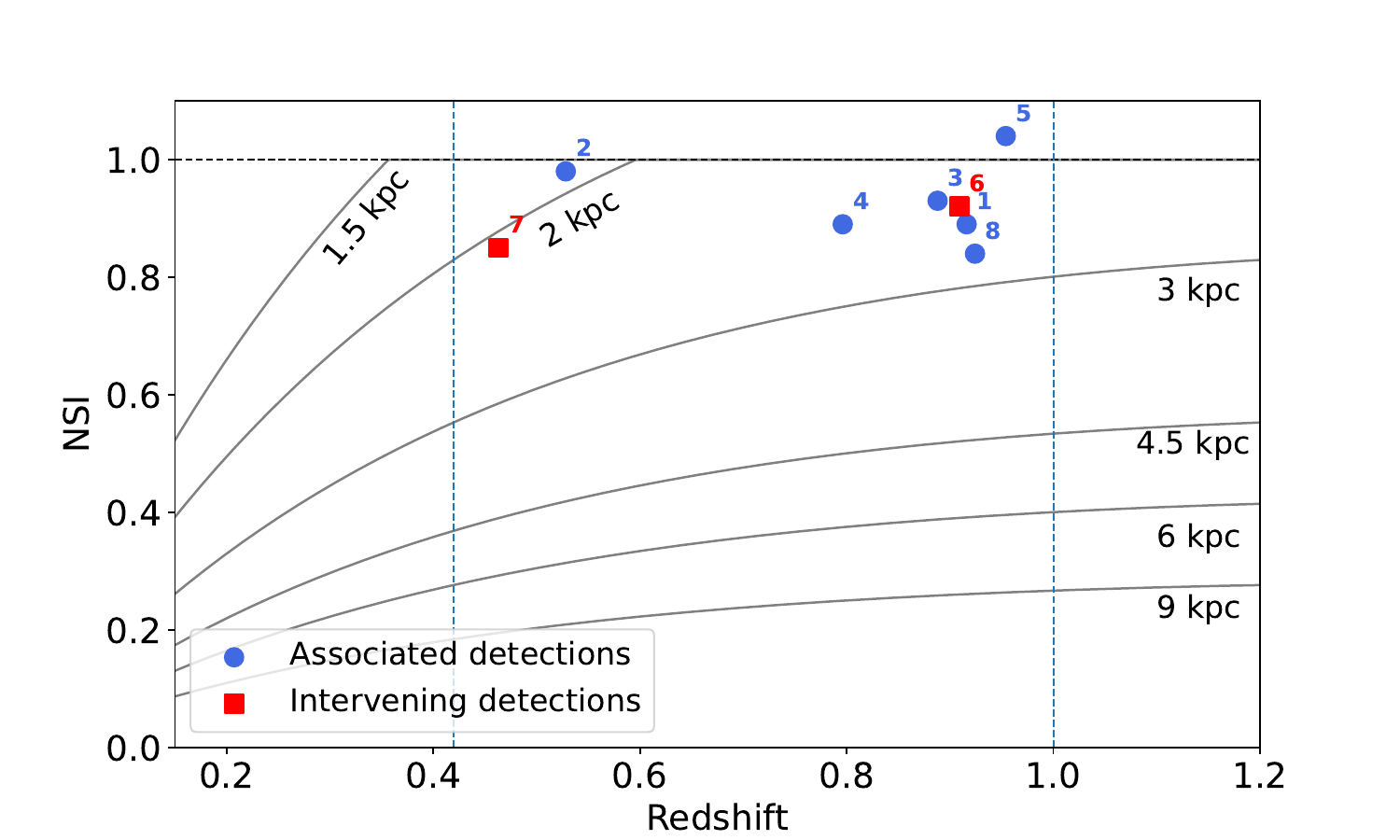}
    \caption{This figure overlays the results for our \hi detections adapted from Figure 6 of \cite{Chhetri2018} for the FLASH survey redshift range $0.42 < z < 1.0$ (blue dotted lines). The intervening systems are shown as red squares and the associated systems in blue filled circles. The numbers correspond to the row numbers in Table \ref{tab:HIdata}.}
    \label{fig:compact_linearsizes}
\end{figure}

Figure \ref{fig:compact_linearsizes} shows the estimated linear sizes of the radio sources with associated \hi lines in Table \ref{tab:HIdata}. This figure is adapted from Fig. 6 of \cite{Chhetri2018}, (using equations 4 and 8; \cite{Chhetri2018}). It assumes an observing frequency of 150\,MHz and an effective scattering distance D = 1 AU. For IPS at 162\,MHz, it uses $\theta_{source} = \theta_{Fresnel} / NSI$, where $\theta_{Fresnel} = 0.3"$.

The IPS NSI values for the radio sources with \hi detections (Figure \ref{fig:compact_linearsizes}) imply a typical projected linear size < 2--3\,kpc, though with some individual scatter.

For associated \hi absorption systems, where the radio source is close to the centre of a galaxy, this is similar to the typical extent of a galactic disk likely to host cold gas. Such a gas disk could provide both the \hi seen in absorption and free-free absorbing material (from the ionised component) that produces a peaked radio SED. 

Simulations (e.g. Young et al. in prep.) show that because jet feedback `spreads out' gas velocities, the deepest \hi absorption is likely to be seen when sources are very compact. Complications to this picture include jets inclined with respect to the disk, and the possibility of bright compact components (hotspots) scintillating even in sources larger than 0.3\,arcsec. Broadly speaking, though, the models suggest that peaked-spectrum sources should evolve towards a power-law SED before their NSI value drops below 0.8. 
The presence of both peaked-spectrum and power-law sources in the NSI-selected sample in Table \ref{tab:WRcrossmatch} is consistent with this picture. 

\subsection{\hi column density estimates}
To estimate the \hi column density of cold gas in these radio sources, we assume an effective spin temperature of 100\,K and covering factor of 1 as is commonly used in previous studies \citep[e.g.][]{Sadler2020}. The 21\,cm \hi absorption line traces only the cool, neutral gas component of gas, so  provides a lower limit to the total gas content \citep{Philstrom2003}. 

We find a median \hi column density of 6.1 $\times$ 10$^{20}$ cm$^{-2}$ for the objects in Table \ref{tab:HIdata}. 
These values are similar to the \hi column densities observed in compact steep-spectrum (CSS) and gigahertz-peaked spectrum (GPS) sources at lower redshift \citep{Chandola2011}. However, it is important to note that these are conservative values, the \hi column density is dependent on the source orientation and morphology. A larger spin temperature would increase the \hi column density, whilst lowering the covering factor decreases the \hi column density. For the detections made, a change in the spin temperature between 100 -- 8000 K and the covering factor between 0.3 -- 1.0, would result in the \hi column density to range between 10$^{20}$ and  10$^{23}$ cm$^{-2}$.

\subsection{Infrared properties from WISE}  
\begin{figure*}
     \centering
             \subfloat[]{\includegraphics[width=0.4\linewidth]{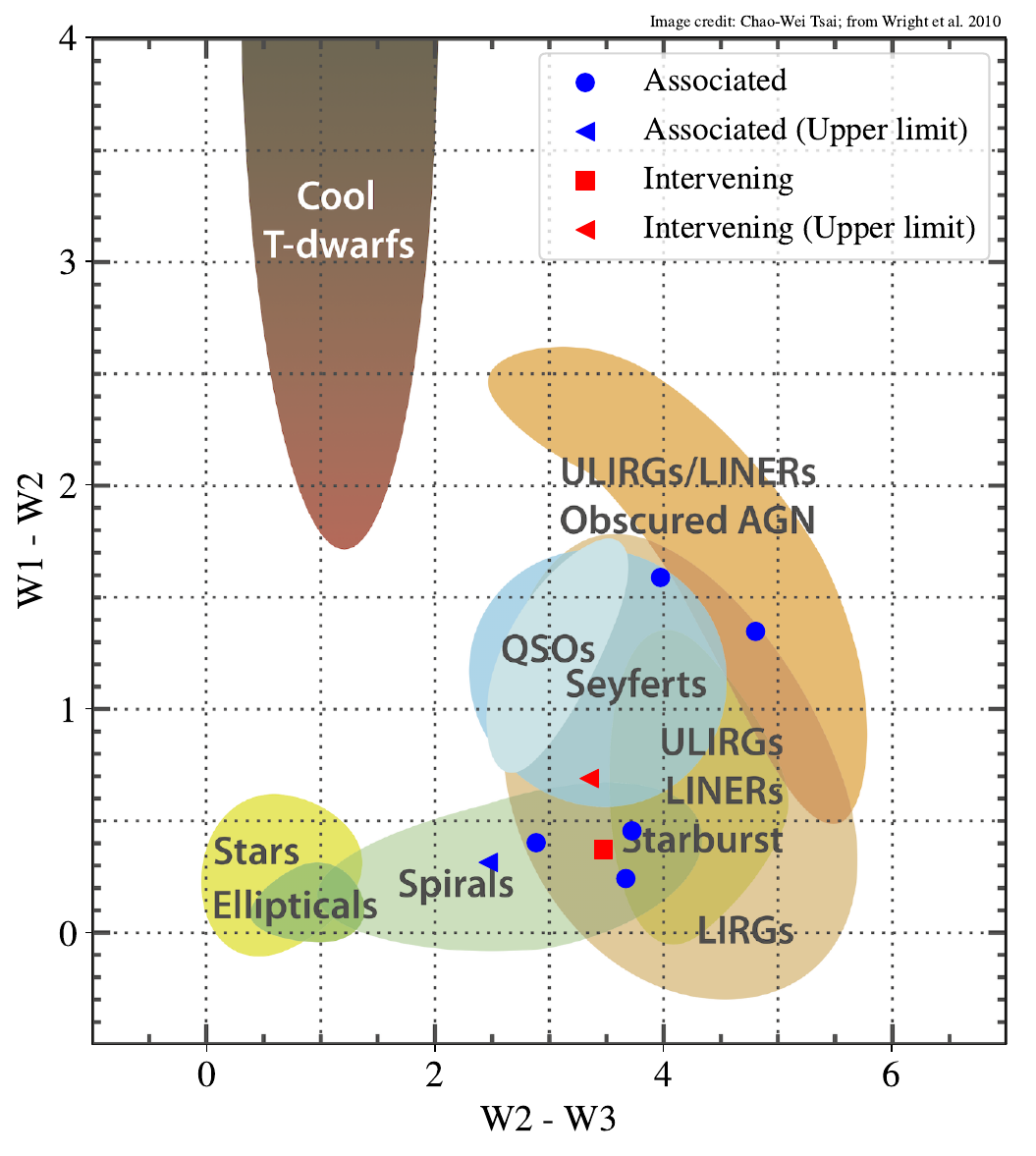}}
                     \subfloat[]{\includegraphics[width=0.4\textwidth]{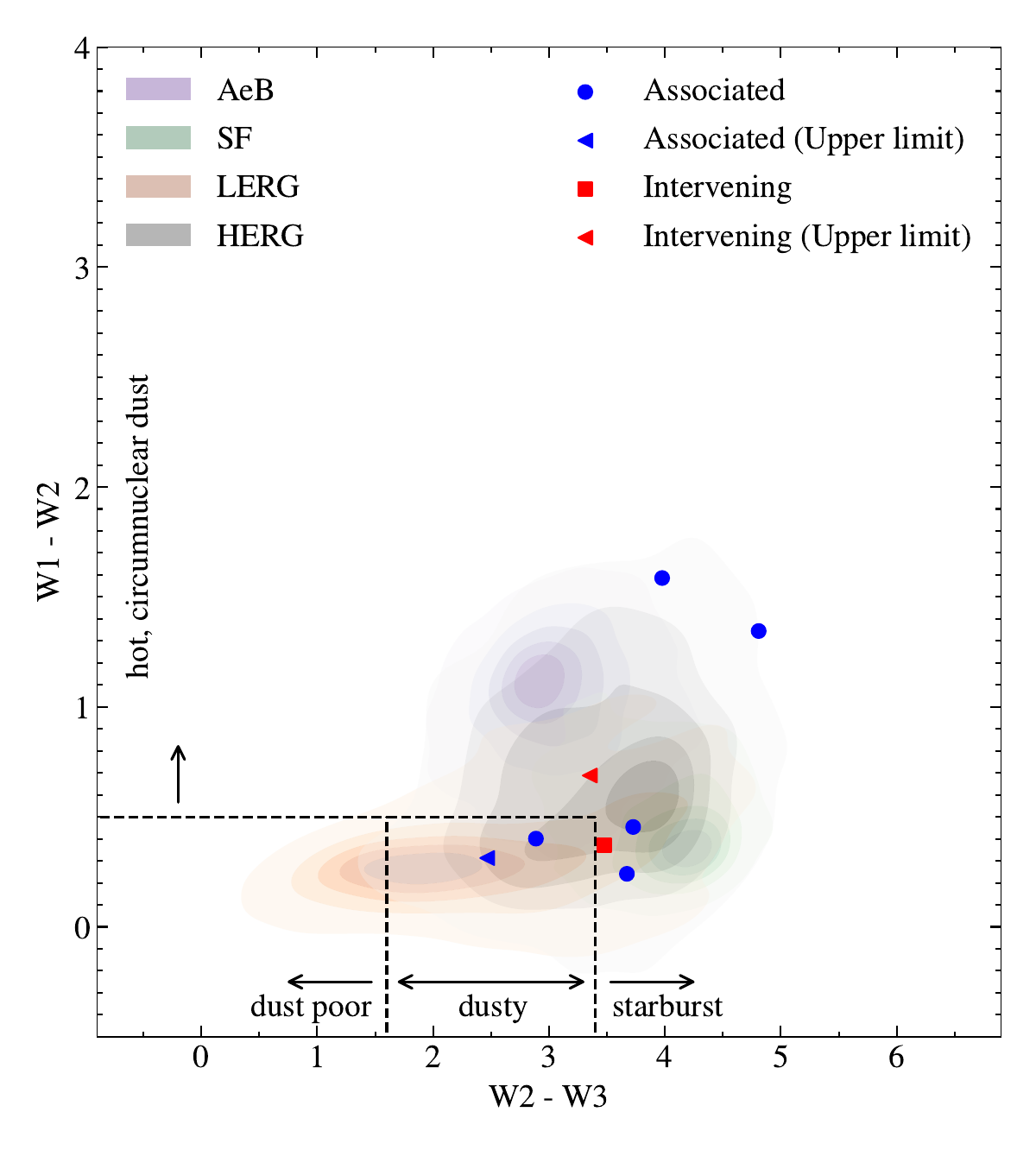}}
     \caption{(a) The \hi detections from the compact radio source sample overlaid on the WISE colour-colour diagram by \cite{Wright2010}. (b) WISE colour-colour diagram using regions from \cite{Ching2017} and \cite{maccagni2017}.
     The potential intervening systems are depicted as red squares as they won't be related to the host source, and the associated systems are presented in blue. The upper limit values illustrated by coloured triangles respectively.}
     \label{fig:wisecolorcolor}
\end{figure*}

Figure \ref{fig:wisecolorcolor}(a) shows the Wide-field Infrared Survey Explorer (WISE) \citep{Cutri2013,Cutri2021} colour-colour diagram from \citet{Wright2010} for the eight radio sources with \hi detections. This diagram is useful to gain an insight on the range of galaxy types present in the \hi detected sample through the relationships between the different WISE infrared bands W1, W2 and W3 (centred at 3.4, 4.6 and 12 micrometres). Crossmatching was conducted using a separation threshold of 2.5 arcsec using the same method as \S2.2. The \hi host galaxies are clearly heterogeneous in nature, spread across the regions covered by spirals, Luminous Infrared Galaxies (LIRGS), starburst \citep{Weedman1981}, Ultra-Luminous Infrared Galaxies (ULIRGS), Low-Ionisation Nuclear Emission Line Regions (LINERs) \citep{Heckman1980}, Quasi-Stellar Objects (QSOs) \citep{Hoyle1966,Urry1995} and Seyferts \citep{Seyfert1943,Peterson2008,Wright2010}. 

Figure \ref{fig:wisecolorcolor}(a) shows that the radio sources with associated \hi absorption are mainly in host galaxies with intermediate to high star formation rates (W2 - W3 $>$ 2). Only two of the six associated lines are in AGN dominated hosts (W1 - W2 $>$ 0.8). Interestingly, there are no sources that are associated with QSOs or Seyferts, most likely due to ionisation of neutral hydrogen by ultraviolet radiation in these objects \citep{Bahcall1969,Curran2012}.

\cite{Ching2017,maccagni2017} further classified various populations of host galaxies at comparable redshifts. In general, the WISE properties of the host galaxies of associated \hi lines in our sample appear similar to those of the general population of high-excitation radio galaxies (HERGs) in this redshift range, \citep[see e.g. Fig.\ 15 of ][]{Ching2017}. Figure \ref{fig:wisecolorcolor}(b), uses \cite{Ching2017} and \cite{maccagni2017} to highlight that the \hi detections occupy regions of dusty star-forming, low-excitation radio galaxies (LERGs) and HERGs whilst no detections are present in the broad emission-line (AeB) objects region consistent with Figure \ref{fig:wisecolorcolor}(a).

\section{Discussion}
\label{sec:discussion}
The initial aim of this study was to use the technique of interplanetary scintillation to compile a sample of compact sources that could be searched for \hi absorption lines in the FLASH redshift range of $0.42 < z < 1.0$.  

Our IPS compact-source sample contains 268 radio sources, 94 of which were estimated to lie within the redshift range $0.42<z<1.0$ where FLASH can detect \hi absorption. Eight \hi lines were detected, all on sightlines to peaked-spectrum sources. No significant detections were found against compact power-law sources, despite their inclusion in the parent sample. 
We now discuss the properties of the \hi-detected objects within the broader context of their parent sample of compact sources. 
 
\subsection{Selection Effects and Biases}
\label{sec:selection effects}
Even without biases, considering the eight \hi detections all associated with peaked spectra, the small sample size would not be statistically significant. However, several selection effects must be considered when interpreting the results of our \hi search. 

As we have already noted, the FLASH survey is sensitive only within the redshift range $0.42 < z < 1.0$ and \hi outside this range is not probed here. 

\subsubsection{Flux density effects}
Even within the FLASH redshift range, \hi absorption can only be detected toward sufficiently bright radio continuum sources. Sources below the continuum sensitivity threshold will not yield detectable absorption even if neutral gas is present.
In a survey like FLASH, where all sources are observed for the same integration time, lines with a wide range in peak optical depth $\tau_{\rm pk}$ can be detected against the brightest sources in each field, but only lines with high optical depth can be detected against fainter sources. 

\begin{figure}[h]
    \centering
    \includegraphics[width=0.95\linewidth]{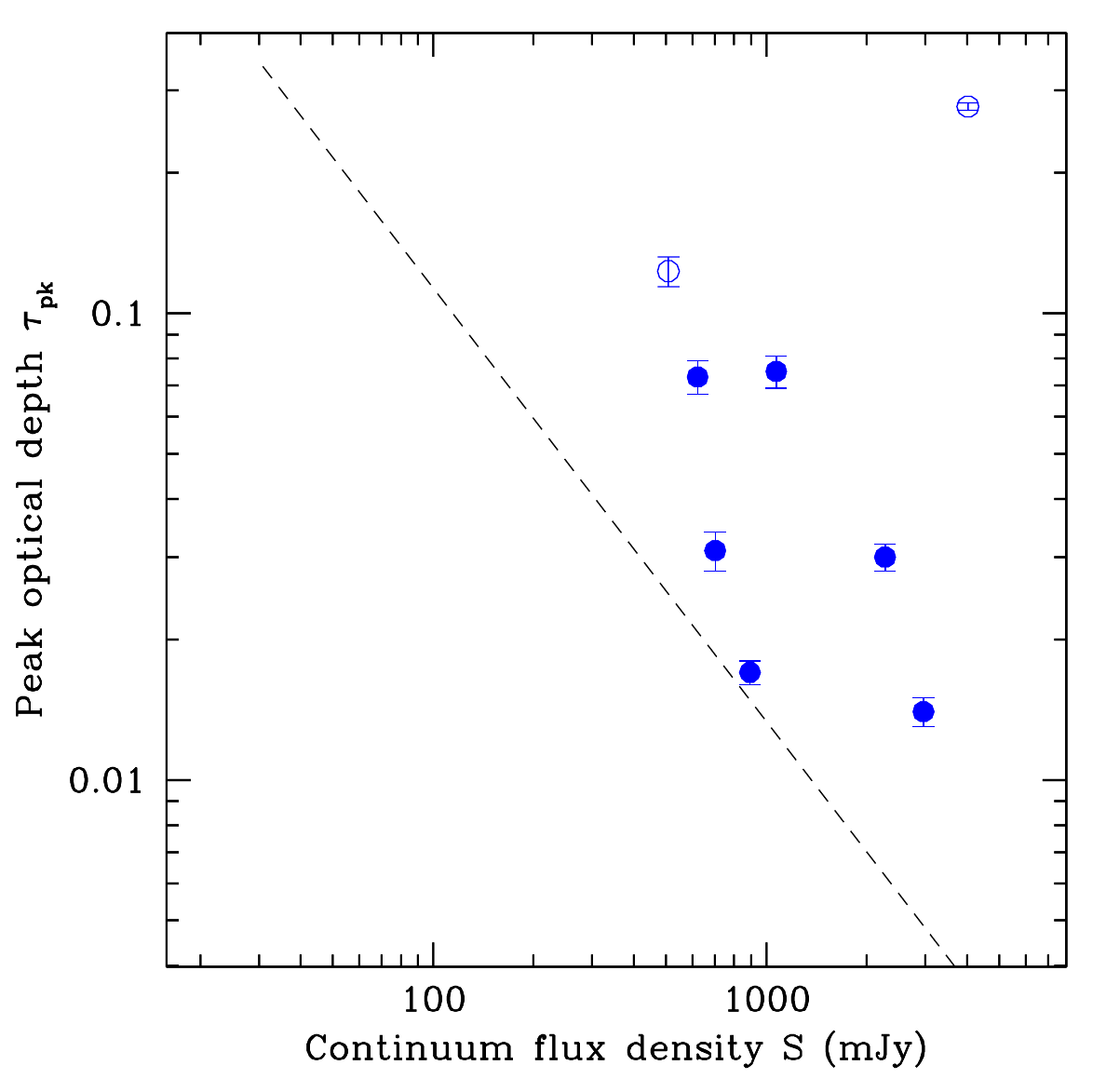}
    \caption{The dashed line shows the indicative 3$\sigma$ detection limit in \hi peak optical depth $\tau_{\rm pk}$ for compact FLASH sources across a range in continuum flux density. Filled blue circles show detections of associated \hi absorption lines, while open circles show the two lines identified as intervening in \S5.3.  } 
    \label{fig:samhita4}
\end{figure}
 
To estimate the detection limit for compact sources spanning a range of flux densities, we adopt the methodology of  \cite{Su2022}, and set an upper limit to the peak optical depth for each source by assuming: 

\begin{equation}\label{equation:column_density_optically_thin}
\begin{split}
\tau_{\rm pk,limit} = \ln\left(\frac{S_{\rm cont}}{S_{\rm cont}-3\sigma_{\rm spec}}\right)
\end{split}
\end{equation}

where S$_{\rm cont}$ is the continuum flux density in mJy and $\sigma_{\rm spec}$ is the rms spectral line noise in a single channel. We use a single value of  $\sigma_{\rm spec}=5.5$\,mJy\,beam$^{-1}$\,channel$^{-1}$ for all spectra. 

Figure \ref{fig:samhita4} shows the indicative detection limit in \hi optical depth $\tau_{\rm pk,limit}$ across the 856\,MHz flux density range spanned by our compact source sample. The eight detected lines are also plotted. It is notable that all eight detections are on sightlines to sources with 856\,MHz flux densities above 500\,mJy, which make up less than 30\% of the IPS compact sample.  

\subsubsection{Peaked-spectrum and power-law sources}
We now consider the 856\,MHz flux density distributions of peaked-spectrum and power-law sources within the redshift range where FLASH can detect associated \hi absorption. 

\begin{figure}[h]
    \centering
    \includegraphics[width=0.98\linewidth]{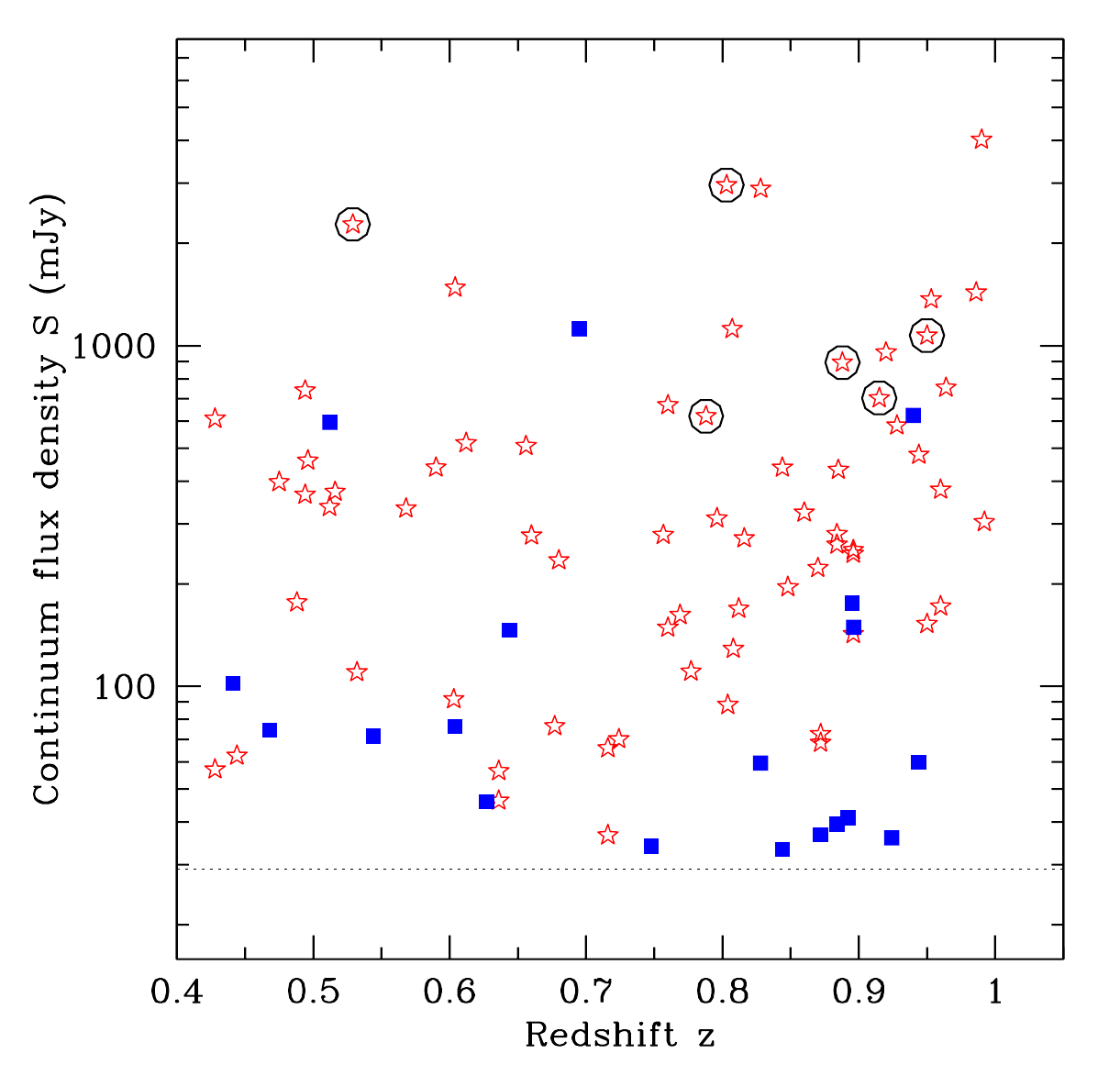}
    \caption{The peak continuum flux density at 856\,MHz against redshift for sources that lie within the redshift range $0.42<z<1.0$ where FLASH can detect associated \hi absorption. Peaked-spectrum sources are shown by open red stars, and sources with power-law spectra by blue filled squares. Black open circles show the location of the six objects with a detection of associated \hi absorption.}
    \label{fig:samhita2}
\end{figure}

Figure \ref{fig:samhita2} shows the distribution of peak flux densities as a function of redshift for sources with optical redshifts within the FLASH band.  There is a clear difference between the peaked-spectrum (PS) and power-law (PL) objects in this redshift range, with a median flux density of 304.0\,mJy for the PS objects and 71.5\,mJy for the PL sources. The brightest sources in Figure \ref{fig:samhita2}, with flux densities above 500\,mJy, have radio luminosities above 10$^{27}$\,W\,Hz$^{-1}$. These are extremely powerful radio AGN, and it is notable that peaked-spectrum sources completely dominate the sample at these high flux densities (over 90\% of the sources brighter than 300\,mJy have peaked radio SEDs). 

The differences in flux density distribution for PS and PL sources means that we cannot make a direct comparison between the \hi detection rates for the two classes, simply because the relative faintness of the PL sources means that they are less sensitive as probes of \hi absorption. 

\begin{figure}[h]
    \centering
    \includegraphics[width=0.98\linewidth]{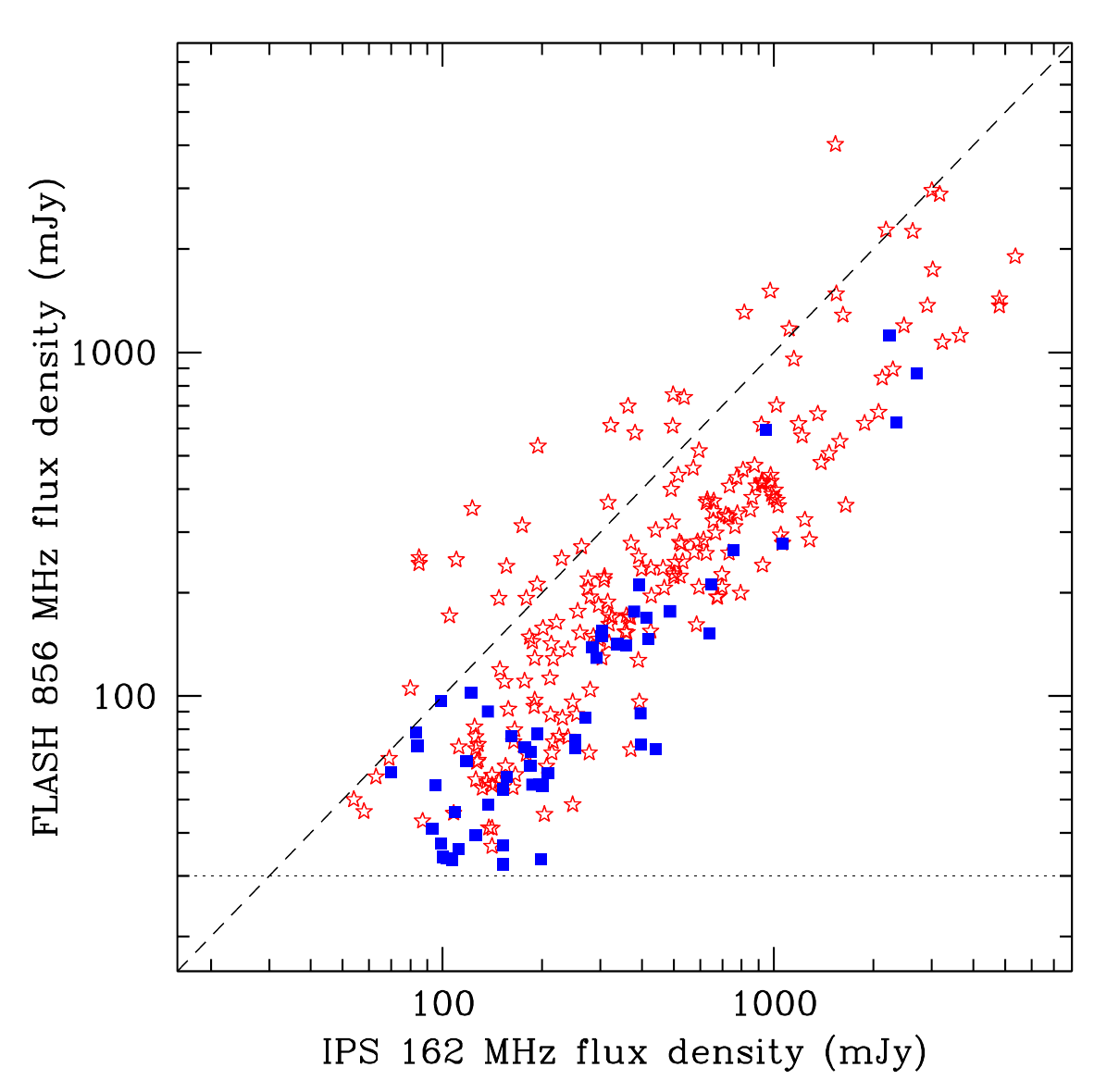}
    \caption{Comparison of 162\,MHz flux densities from the IPS catalogue and 856\,MHZ flux densities from FLASH for the compact-source sample from Table \ref{tab:WRcrossmatch}. As in Figure \ref{fig:samhita2}, PS sources are shown by red stars and PL sources by blue squares. The horizontal dotted line marks the 30\,mJy cutoff for the FLASH sample. The diagonal dashed line corresponds to the spectral index line of 0.}
    \label{fig:samhita5}
\end{figure}

The IPS selection method for compact sources makes no distinction between objects with peaked and power-law radio spectra. However, the MWA IPS measurements are made at 162\,MHz while the FLASH flux densities plotted in Figure \ref{fig:samhita2}
are at 856\,MHz. Sources with a power-law spectrum will generally be fainter at 856\,MHz than at 162\,MHz, while the opposite might be true of peaked-spectrum sources if the peak lies above 162\,MHz. 
To check this, we compared the flux densities of the sources in Table \ref{tab:WRcrossmatch} at 162\,MHz (IPS) and 856\,MHz (FLASH) as shown in Figure \ref{fig:samhita5}. As can be seen from this figure, peaked-spectrum sources are also the dominant population among the brightest sources at 162\,MHz, consistent with the earlier findings of \cite{Chhetri2018}. 
 
We can also see from Figure \ref{fig:samhita5} that compact sources in the MWA IPS sample typically have 162\,MHz flux densities above 80--100\,mJy, corresponding to the sensitivity limit for IPS detections in the MWA search. As a result, the sample of IPS-selected FLASH sources is also skewed towards higher 856\,MHz flux densities than the overall FLASH sample studied by \cite{yoon2025}. \\

\noindent It can be noted, no \hi detections have been detected towards sources classified with a retriggered spectrum. However, the sample size of 10 sources with retriggered SEDs does not provide a statistically robust assessment of whether there is a physical mechanism behind their null result.

\subsection{Detection rate for associated \hi absorption}
\label{sec:detrate} As discussed in \S\ref{sec:props} and summarised in Table \ref{tab:opt}, 94 of our 268 sources have measured or estimated optical redshifts in the range $0.42<z<1.0$ where associated \hi absorption could be detected in FLASH spectra. 

The detection rate for an \hi absorption survey will depend on the sensitivity of the observations and the optical depth limit reached. 
For example, \cite{vermeulen2003} found a detection rate of 33\% (19$/$57 sources) for associated \hi absorption in a targeted sample of bright compact radio sources but their detections included lines with peak optical depths as low as $\tau_{\rm pk}<$ 0.001 which would not be detectable by FLASH. Another intermediate redshift \hi absorption survey targetting compact radio sources by \cite{Grasha2019}, found a lower detection rate of 5.6\% with $\tau_{\rm pk}$ down to 0.0015. To compare detection rates between different surveys and subsamples, we therefore need to measure the detection rate above some fixed value of $\tau_{\rm pk}$ \citep{Su2022}. 

For the sample in Table \ref{tab:WRcrossmatch}, 
we set a value of $\tau_{\rm pk}\geq$ 0.03 for the detection rate measurement. This value was chosen as a compromise between sample size and sensitivity - for lower limiting values of $\tau_{\rm pk}$ the number of sources bright enough to make a detection decreases, while for higher values more sources can be included but the number of detections strong enough for inclusion in the sample decreases. 

Of the 94 sources in the redshift range $0.42<z<1.0$ where FLASH could detect associated absorption, 25 are bright enough to detect a line with $\tau_{\rm pk}\geq$ 0.03 and there are four associated \hi lines detected above this optical-depth limit. The \hi detection rate for $\tau_{\rm pk}\geq$ 0.03 is therefore 16 $\pm$ 8\%. We can derive a comparable rate for the low-redshift sample of compact sources searched by \cite{maccagni2017}, for which the detection rate at $\tau_{\rm pk}\geq$ 0.03 is 9$/$39 or 23 $\pm$ 7\%. Although our lower detection rate is consistent with a general decrease in the \hi detection rate with increasing redshift \citep{aditya2016,Grasha2019,Su2022}, the difference in \hi detection rates between our sample (median z $=$ 0.83) and that of \cite{maccagni2017} (median z $=$ 0.08) and also when applying the optical depth limit to \cite{Grasha2019} (with a median z of $\sim$0.6) is not statistically significant based on the size of the samples.

\subsection{Physical Interpretation}
Peaked-spectrum radio sources are widely interpreted as young radio galaxies in which jets are still expanding within the dense interstellar medium of the host galaxy \citep{O'Dea2021}.  In this phase, the radio emission region remains compact (1--3\,kpc in size) and the covering factor of foreground neutral gas relative to the background continuum source is likely to remain high. A higher covering factor naturally enhances the probability of detecting associated \hi absorption. In addition, a dense circumnuclear environment may both confine the radio source and provide the neutral gas responsible for the observed \hi absorption. Given a source size, the age and spectrum of the source depends on the environment into which the jets propagate \cite{Bicknell2018,Young2025}, such as in the alternative `frustrated' interpretation.

The detection of \hi absorption in individual sources in the FLASH survey is likely to depend on several factors, including the continuum flux density, the gas temperature and density distribution, and the orientation of the radio jet with respect to the observer. As a result, the 16\% detection rate for associated \hi absorption derived in \S\ref{sec:detrate} is probably a lower limit to the fraction of compact radio sources that contain cold and dense circumnuclear gas. 

It is interesting to compare the associated \hi absorption detections in Table \ref{tab:HIdata} with the models for young radio galaxies recently published by \cite{Young2025}. 
We detect \hi absorption in compact, peaked-spectrum sources with typical linear sizes of 2--3\,kpc and 856 MHz luminosities above about $10^{27}$\,W\,Hz$^{-1}$. 
These parameters correspond reasonably well with the Young et al.\ Q44 jet models, which have a radio jet power of $10^{44}$\,erg\,s$^{-1}$ and jet ages of around 0.2 to 0.4\,Myr for a source 2--3\,kpc in extent. 

Our results appear to support an evolutionary framework in which radio sources begin in a compact, peaked-spectrum phase. During this early phase \citep[lasting up to 1\,Myr;][] {Kukreti2024}, they are embedded in dense neutral gas. As jets expand beyond the host galaxy, the radio spectrum evolves toward a power-law form and the probability of detecting associated \hi absorption decreases.

If correct, this scenario implies that the peaked-spectrum phase represents a critical stage for studying AGN fuelling and feedback.

\section{Conclusions and future work}
\label{sec:conclusion}
We have combined interplanetary scintillation measurements from the MWA with ASKAP FLASH survey data to construct a sample of compact (size $<$ 2--3\,kpc) radio sources in the redshift range $0.42 < z < 1.0$. Using Bayesian RadioSED modelling, we classified their radio spectral energy distributions and searched for associated \hi absorption.

Our main conclusions are as follows:
\vspace*{-0.2cm}
\begin{itemize}
\item	
The compact extragalactic sources identified through IPS contain a high fraction of peaked-spectrum systems, as also found by earlier IPS studies \citep{Chhetri2018,Sadler2019}. 
\item
All eight of the \hi-detected sources have peaked radio spectra. It is not yet clear whether peaked-spectrum sources are more gas-rich than other compact sources, since our sample contained too few bright `non-peaked' sources to provide a statistical test. 
\item	
The detected \hi systems all have estimated \hi column densities above $2\times10^{20}$\,cm$^{-2}$. This result supports the hypothesis that peaked-spectrum sources coincide with substantial neutral gas reservoirs in the host galaxies, although this result is not statistically definitive due to the small sample size.
\item 
Observationally, Very Long Baseline Interferometry (VLBI) imaging would provide higher resolution for the \hi-detected sources, offering valuable information about the source morphology. When compared with high-resolution simulations of jet-gas interactions (like those of \citet{Bicknell2018} and \citet{Young2025}), this may offer powerful insights into the effects of the neutral gas properties and source geometry on the observed \hi absorption profile. 
\end{itemize}

This work is a preliminary study of the full FLASH survey using approximately 12\% of the data that will be available at the completion of the full survey. Future work will expand the sample using the full FLASH survey and incorporate additional multi-wavelength data to further investigate host galaxy properties and the role of environment in shaping radio source evolution. 

\begin{acknowledgement}
This scientific work uses data obtained from Inyarrimanha Ilgari Bundara, the CSIRO Murchison Radio-astronomy Observatory. We acknowledge the Wajarri Yamaji People as the Traditional Owners and native title holders of the Observatory site. CSIRO’s ASKAP radio telescope is part of the Australia Telescope National Facility (https://ror.org/05qajvd42). Operation of ASKAP is funded by the Australian Government with support from the National Collaborative Research Infrastructure Strategy. ASKAP uses the resources of the Pawsey Supercomputing Research Centre. Establishment of ASKAP, Inyarrimanha Ilgari Bundara, the CSIRO Murchison Radio-astronomy Observatory and the Pawsey Supercomputing Research Centre are initiatives of the Australian Government, with support from the Government of Western Australia and the Science and Industry Endowment Fund.

This work was supported by the Australian SKA Regional Centre (AusSRC), Australia’s portion of the international SKA Regional Centre Network (SRCNet), funded by the Australian Government through the Department of Industry, Science, and Resources (DISR; grant SKARC000001). AusSRC is an equal collaboration between CSIRO – Australia’s national science agency, Curtin University, the Pawsey Supercomputing Research Centre, and the University of Western Australia.

This research has made use of the NASA/IPAC Extragalactic Database (NED), which is operated by the Jet Propulsion Laboratory, California Institute of Technology, under contract with the National Aeronautics and Space Administration.

HY is supported by the National Research Foundation of Korea (NRF) grant funded by the Korea government (MSIT, RS-2025-00516062), and by funding from Korean government (Korea AeroSpace Administration (KASA), grant number RS-2026-25587698).

MG is supported through UK STFC Grant ST/Y001117/1. MG acknowledges support from the Inter-University Institute for Data Intensive Astronomy (IDIA). IDIA is a partnership of the University of Cape Town, the University of Pretoria and the University of the Western Cape. For the purpose of open access, the author has applied a Creative Commons Attribution (CC BY) licence to any Author Accepted Manuscript version arising from this submission.

RS acknowledges the support from the National Science Foundation of China (grant 12603034), the Shanghai Super Postdoctoral Incentive Program (No. 2025322), the China Postdoctoral Science Foundation (Grant No. 2025M783232) and the National SKA Program of China (No. 2025SKA0130100).

We also thank the anonymous referee for their constructive and helpful feedback and comments for our manuscript.
 
\end{acknowledgement}

\bibliography{bibliography}

\begin{landscape}
\begin{table}[p]
    \centering
\caption{The first twenty rows of the table of 268 compact sources searched for \hi absorption. The columns listed are as follows: (1) Source name, (2) FLASH catalogue name, (3) Observation field number (SBID) and island number from the CASDA database, (4, 5) Radio source RA and Dec (J2000), (6, 7) Peak and total continuum flux density at 856\,MHz, (8) Detection limit in \hi optical depth $\tau_{\rm lim}$ for a 3-$\sigma$ detection in a single spectral channel, (9, 10) IPS Normalised Scintillation Index (NSI) at 162\,MHz, and its associated uncertainty, as discussed in \S2.2, (11) Radio SED classification, as discussed in \S\ref{sec:radioSED}, (12) Optical classification (as galaxy, quasar, or blank field), (13) LS10 optical type \cite{Dey2019}, PSF: Point source, SER: Sersic profile, REX: round exponential galaxies with a variable radius, DEV: deVaucouleurs profile, EXP: exponential profile, (14, 15) Optical redshift, and whether this comes from spectroscopic (spec) or photometric (phot) data. Spectroscopic and photometric redshifts are given to three and two significant figures respectively, (16) Source of the optical redshift: For photometric redshifts, R26: \cite{Roster2026}, Z21: \cite{zhou2021}. For spectroscopic redshifts, deV07: \cite{devries2007}, Br75: \cite{browne1975}, Bu72: \cite{burbidge1972}, DR1: \cite{DESI2026}, Dr97: \cite{Drinkwater1997}, Hu78: \cite{hunstead1978}, Ja02: \cite{jackson2002}, Ma11:\cite{mahony2011}, Mc96: \cite{mccarthy1996}, Mo16: \cite{monroe2016}, St94: \cite{stickel1994}, Wr79: \cite{wright1979}. } 
\label{tab:WRcrossmatch}
   \begin{tabular}{lllrrrrllllllrllll}

    \hline
        NED name & FLASH name & FLASH SBID & RA & Dec & S$_{\rm peak}$ &  S$_{\rm int}$ & $\tau_{\rm pk,limit}$ & NSI & $\pm$ & SED class & Optical ID & LS10 type & z & z-type & z\_ref \\ 
     & & & deg & deg & mJy & mJy \\
     (1) & (2) & (3) & (4) & (5) & (6) & (7) & (8)& (9) & (10) & (11) & (12) & (13) & (14) & (15) & (16) \\  
        \hline
        PKS 0019+058  & J002232+060803 & SB63396\_20 & 5.635174 & 6.134247 & 351.3 & 362.7 & 0.048 & 0.88 & 0.05 & peaked & Quasar & PSF & >  2.86 & spec & deV07 \\ 
        NVSS  J002249-034425 & J002249-034424 & SB63446\_174 & 5.706597 & -3.740230 & 72.5 & 73.9 & 0.258 & 0.91 & 0.08 & peaked & Galaxy & SER & 0.87 & phot & R26  \\ 
        NVSS  J002903+050934  & J002903+050933 & SB63396\_12 & 7.265211 & 5.159288 & 501.6 & 525.8 & 0.033 & 0.88 & 0.05 & retriggered & Quasar & PSF & 1.640 & spec & DR1  \\ 
        NVSS  J003157+025520 & J003157+025519 & SB63396\_48 & 7.991209 & 2.922186 & 199.4 & 208.1 & 0.086 & 0.91 & 0.05 & peaked & Galaxy  & REX & 1.43 & phot & R26  \\ 
        PKS 0029-09 & J003200-085136 & SB63446\_5 & 8.003864 & -8.860088 & 843.9 & 886.3 & 0.020 & 0.83 & 0.04 & peaked & Faint/blank & none & > 1 & phot & .. \\
        &&&& \\
        NVSS J003253+065023 & J003254+065022 & SB63396\_232 & 8.225227 & 6.839685 & 55.1 & 57.7 & 0.356 & 0.92 & 0.08 & peaked & Faint/blank & none & > 1 & phot & ..  \\ 
        NVSS  J003455-071159 & J003455-071200 & SB63446\_237 & 8.729225 & -7.199793 & 54.0 & 56.2 & 0.364 & 1.02 & 0.15 & peaked & Faint/blank & none & > 1 & phot & ..  \\
        NVSS J003657+071243 & J003657+071241 & SB63396\_141 & 9.241181 & 7.211625 & 86.5 & 90.6 & 0.212 & 1.02 & 0.07 & peaked & Faint/blank & none & > 1 & phot & .. \\ 
        NVSS J003803+043651 & J003803+043650 & SB63396\_28 & 9.514477 & 4.614052 & 277.6 & 290.2 & 0.061 & 0.84 & 0.04 & power-law & Faint/blank & none & > 1 & phot & .. \\ 
        NVSS  J003817+054823 & J003817+054821 & SB63396\_78 & 9.572455 & 5.806080 & 141.5 & 148.0 & 0.124 & 0.92 & 0.05 & power-law & Faint/blank & none & > 1 & phot & ..  \\
        &&&& \\
        NVSS  J003820-032959 & J003820-032957 & SB63446\_35 & 9.586584 & -3.499384 & 238.8 & 248.1 & 0.072 & 0.82 & 0.13 & peaked & Quasar & PSF & 1.858 & spec & Ma11 & \\ 
        NVSS  J003836+070705 & J003837+070703 & SB63396\_30 & 9.654362 & 7.117726 & 264.8 & 277.2 & 0.064 & 0.82 & 0.04 & power-law & Galaxy & REX & 1.14 & phot & R26 \\ 
        NVSS  J004158-180451 & J004157-180450 & SB63394\_34 & 10.491354 & -18.080624 & 206.4 & 221.0 & 0.083 & 0.85 & 0.05 & peaked & Galaxy & DEV & 1.05 & phot & R26 \\ 
        NVSS  J004447-213930 & J004447-213929 & SB63394\_67 & 11.197538 & -21.658313 & 136.6 & 145.0 & 0.129 & 0.84 & 0.21 & unsure & Galaxy & DEV & 0.88 & phot & R26  \\ 
        NVSS  J004451-170933 & J004451-170933 & SB63394\_276 & 11.213011 & -17.159358 & 41.2 & 44.4 & 0.512 & 0.89 & 0.21 & power-law & Galaxy & REX & 0.89 & phot & R26  \\ 
        &&&& \\
        NVSS  J004506-071046 & J004506-071045 & SB63446\_30 & 11.276711 & -7.179230 & 277.3 & 287.9 & 0.061 & 0.82 & 0.04 & peaked & Galaxy & DEV & 0.66 & phot & R26  \\ 
         NVSS J004629-215757 & J004629-215757 & SB63394\_318 & 11.621109 & -21.966030 & 35.9 & 38.4 & 0.616 & 0.81 & 0.22 & power-law & Galaxy & DEV & 0.92 & phot & R26 \\ 
         NVSS J004707-182642 & J004707-182643 & SB63394\_337 & 11.782631 & -18.445386 & 33.3 & 35.8 & 0.683 & 0.94 & 0.23 & power-law & Galaxy & REX & 0.84 & phot & R26 \\ 
        NVSS  J004821+054822 & J004821+054821 & SB63396\_131 & 12.090803 & 5.805981 & 91.7 & 96.8 & 0.198 & 0.92 & 0.06 & peaked & Galaxy & SER & 0.603 & spec & DR1  \\ 
        NVSS  J004931-195606 & J004931-195607 & SB63394\_123 & 12.381677 & -19.935340 & 81.4 & 85.9 & 0.226 & 1.02 & 0.21 & peaked & Galaxy & SER & 0.37 & phot & R26  \\ 
        \hline
    \end{tabular}
\end{table}
\end{landscape}

\begin{landscape}
\begin{table}[p]
    \centering
\caption{Table of \hi absorption line detections for the Compact Sample with NSI $\geq$ 0.8. Column (1) Source name, (2) FLASH catalogue name, (3) Observation field number (SBID) and component number from the CASDA database, (4, 5) Radio source RA and Dec (J2000), (6) Peak flux density of the radio source at 856\,MHz (mJy), (7) IPS Normalised Scintillation Index (NSI) at 162 MHz, (8) Radio SED classification, as discussed in \S\ref{sec:radioSED}, (9) SED peak frequency (MHz), (10, 11) Optical redshift, and whether this comes from spectroscopic (spec) or photometric (phot) data. (12) Source of the optical redshift, (13) \hi redshift, (14) Peak optical depth of the \hi line, (15) Integrated optical depth of the \hi line (\kms), (16) Line width (FWHM) of the \hi line (\kms), (17) Log of the Bayes Factor for the significance of the \hi line detection, (18) Tentative line classification as associated (As) or intervening (In). For the optical redshift reference (12): photometric redshifts, R26: \cite{Roster2026}, Z21: \cite{zhou2021}. For spectroscopic redshifts, DR1: \cite{DESI2026}, Mc96: \cite{mccarthy1996}, St94: \cite{stickel1994}. } 
\label{tab:HIdata}
    \begin{tabular}{lllrrrrrrrlrrrrrrrl}
    \toprule
    \hline
 NED name & FLASH Name &  FLASH SBID &  RA &  DEC & S$_{\rm pk}$ & NSI & SED class & $\nu_{\rm peak}$ &  $z_{opt}$ & z-type & z$_{\rm ref}$ & $z_{HI}$ &  $\tau_{pk}$&  $\tau_{int}$ & Width & ln\,(B) & Cl \\
   & & & deg & deg & mJy & & & MHz  & & & && &  \kms & \kms & & \\
  (1) & (2) & (3) & (4) & (5) & (6) & (7) & (8)& (9) & (10) & (11) & (12) & (13) & (14) & (15) & (16) & (17) & (18)\\  
    \hline
 1   MRC 0201-214 &  J020332-211347
&  SB63041 5a &  30.88396&  -21.229939&  702.1 & 0.89 &peaked& 234 & 0.915 & spec & Mc96 & 0.9161 &0.031 &  3.13& 102.7& 50 & As \\
2   PKS 0431-02 & J043355-022957 &  SB63513 2a&  68.479011
&  -2.499367 &  2271.0 & 0.98 &peaked& 459 & 0.529 &spec&  DR1 & 0.5282& 0.030&  1.60& 53.6& 122 & As \\
3   PKS 1057+050 & J110011+044401 &  SB51444 4a & 165.047795 &  4.733664 &  892.8 & 0.93 &peaked & 296 & 0.888 & spec & DR1 & 0.8880&  0.017 &  3.59 & 213.1 & 92 & As \\
4   PKS 1355+01 & J135753+004633 & SB52526 1a & 209.473669
& 0.776038 & 2961.1 & 0.89 &peaked& 504 & 0.80 &phot & Z21 & 0.7962& 0.015 & 1.09& 80.3& 168 & As \\
5   MRC 2125-237 & J212805-232946&  SB55464 2a &  322.023376&  -23.496245&  1070.9 & 1.04&peaked&  87 & 0.950 & spec &  Mc96 & 0.9540&  0.075 &  16.09& 215.9& 126 & As \\
6   PKS 2127+04 & J213032+050215 & SB52547 1a & 322.635949 & 5.037638 & 4027.0 & 0.92 & peaked & 682 & 0.990 & spec & St94 & 0.9088 & 0.277 & 6.15 & 22.2 & 3897 & In \\
7   MRC 2131-241 & J213437-235537&  SB55464 9a &  323.657106 &  -23.927134 &  508.1 & 0.85 &peaked& 148 & 0.66 &phot& R26 & 0.4632&  0.123 &  2.25& 18.2& 198& In \\
 8   NVSS J214834+042137 & J214834+042134 & SB66727 12a &  327.145536 &  4.359537 &  621.5 & 0.84 &peaked& 237 & 0.79 &phot & R26 & 0.9242&  0.073&  3.85& 52.7& 156 & As \\
 \hline
\bottomrule
    \end{tabular}
    
\end{table}
\end{landscape}

\appendix

\clearpage
\onecolumn
\section{Detected HI Absorption Lines}
\label{sec:lines}
\vspace{1cm}

\begin{figure}
     \centering
     \caption{These diagrams show the spectral line profiles of the strong \hi detections with the outputs from the \textit{FLASHfinder}, with the line width in [km/s] centred at 0 against the ratio of the flux density of the line to the continuum flux density. The grey shaded region refers to a 1-$\sigma$ noise value.}
             \subfloat[FLASH ID SB63041 5a]{\includegraphics[width=0.5\textwidth]{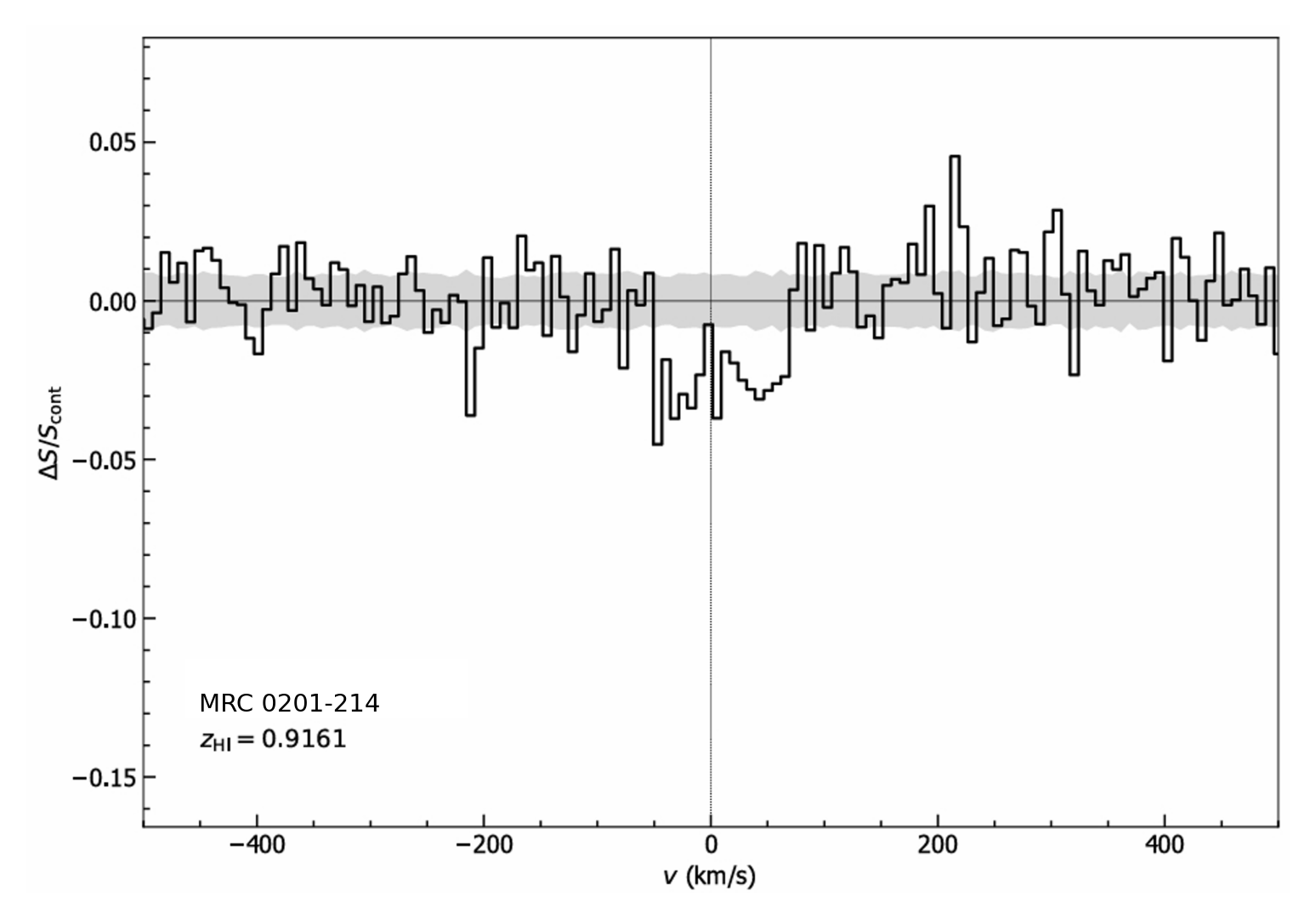}}
                     \subfloat[FLASH ID SB63513 2a]{\includegraphics[width=0.5\textwidth]{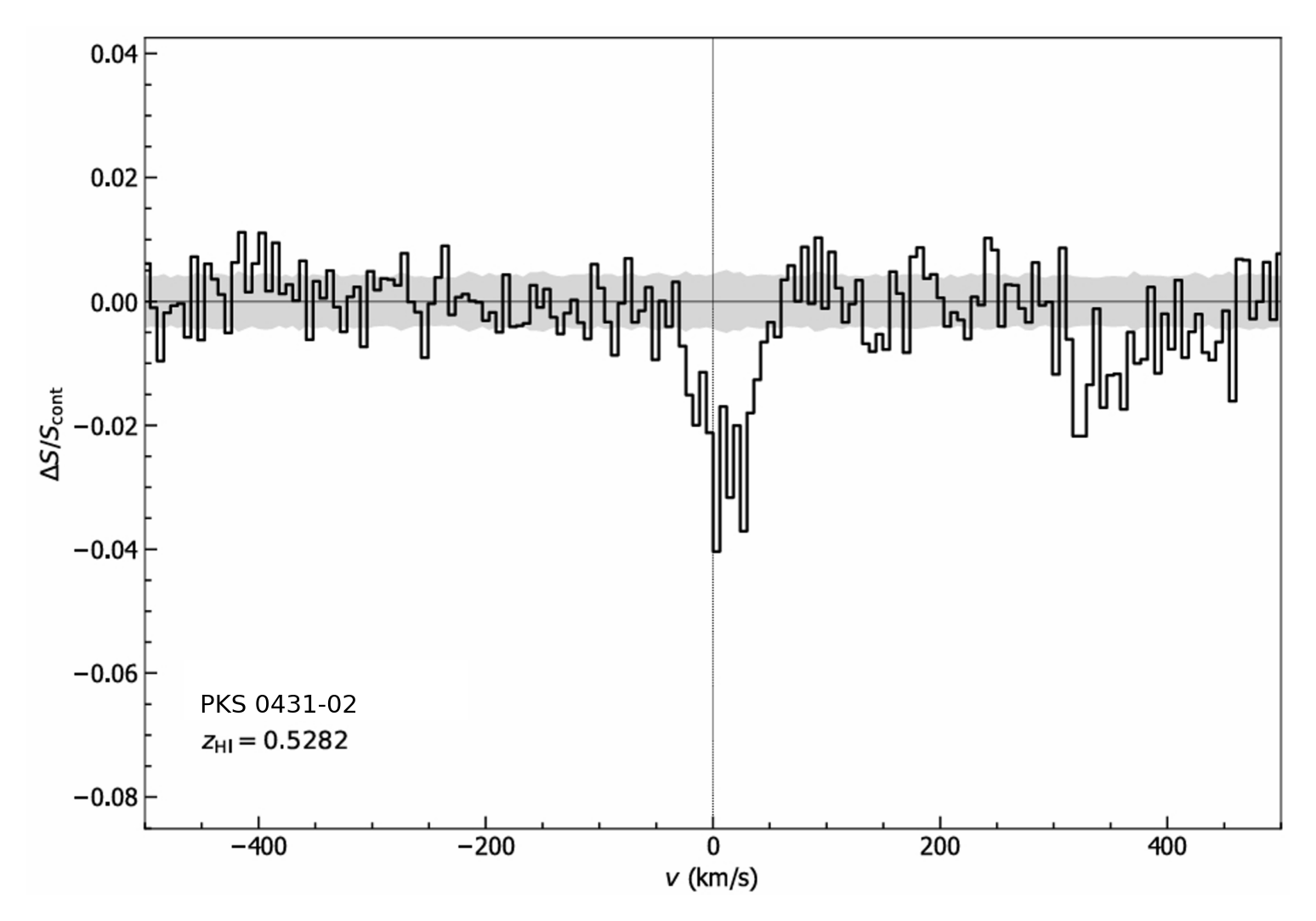}}\\
    \vspace{1.5cm}
     \subfloat[FLASH ID SB51444 4a]{\includegraphics[width=0.5\textwidth]{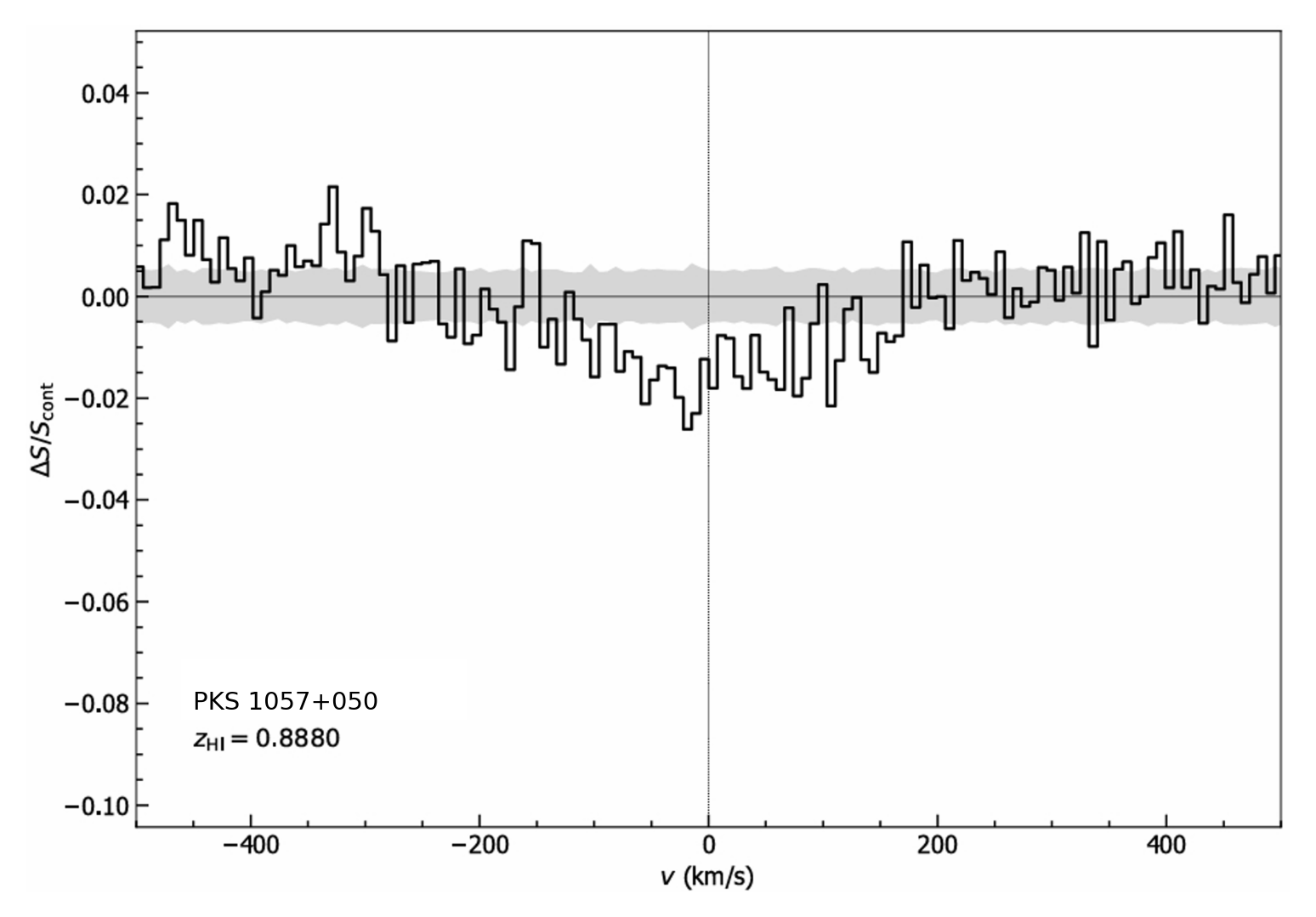}}
        \subfloat[FLASH ID SB52526 1a]{\includegraphics[width=0.5\textwidth]{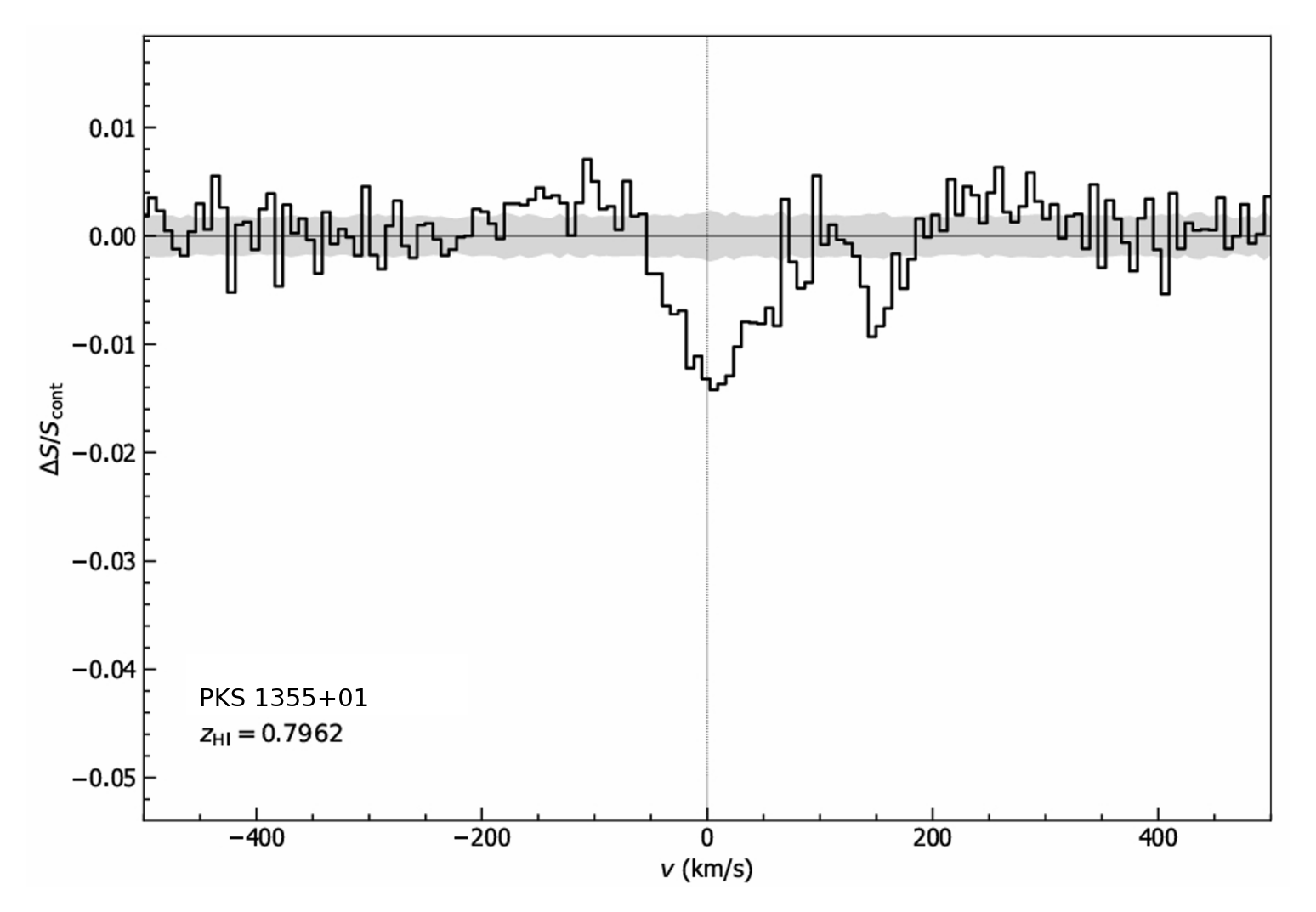}}
\end{figure}
\newpage
\vspace{1cm}

        \begin{figure}
        \centering
        \ContinuedFloat
        \caption{Spectral line profiles continued.}
                \subfloat[FLASH ID SB55464 2a]{\includegraphics[width=0.5\textwidth]{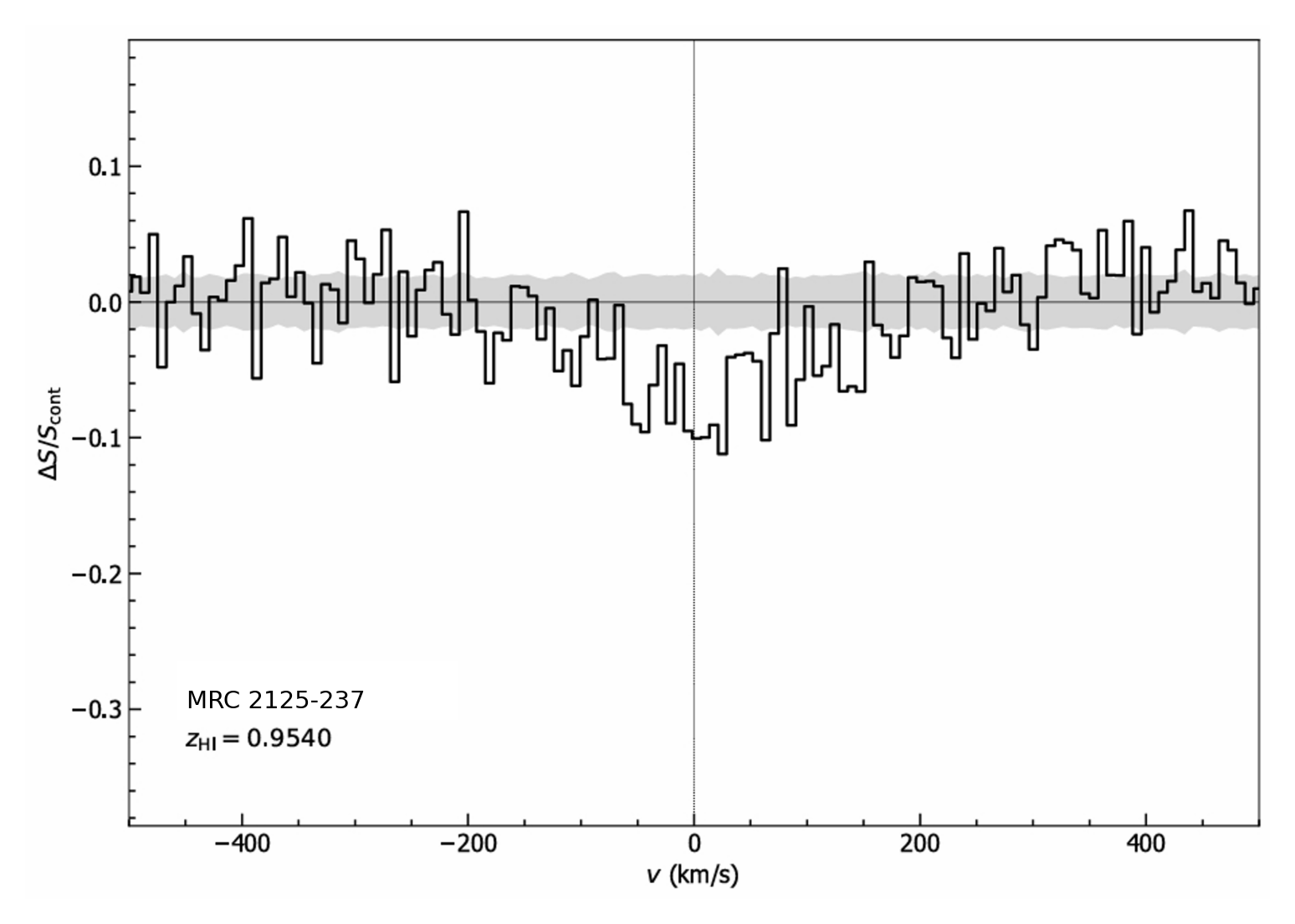}}
        \subfloat[FLASH ID SB52547 1a]{\includegraphics[width=0.5\textwidth]{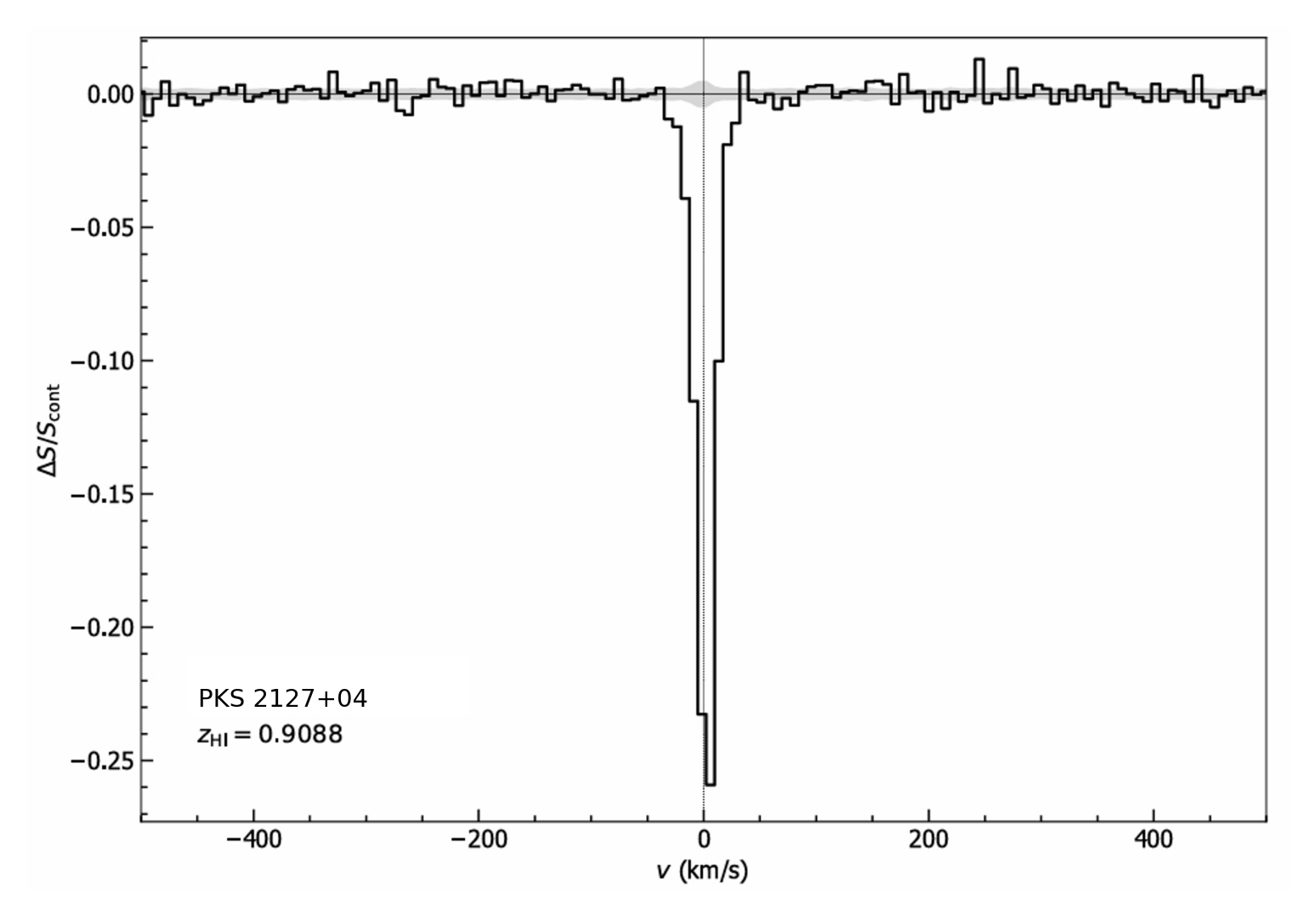}}\\
                \vspace{1.5cm}
                \subfloat[FLASH ID SB55464 9a]{\includegraphics[width=0.5\textwidth]{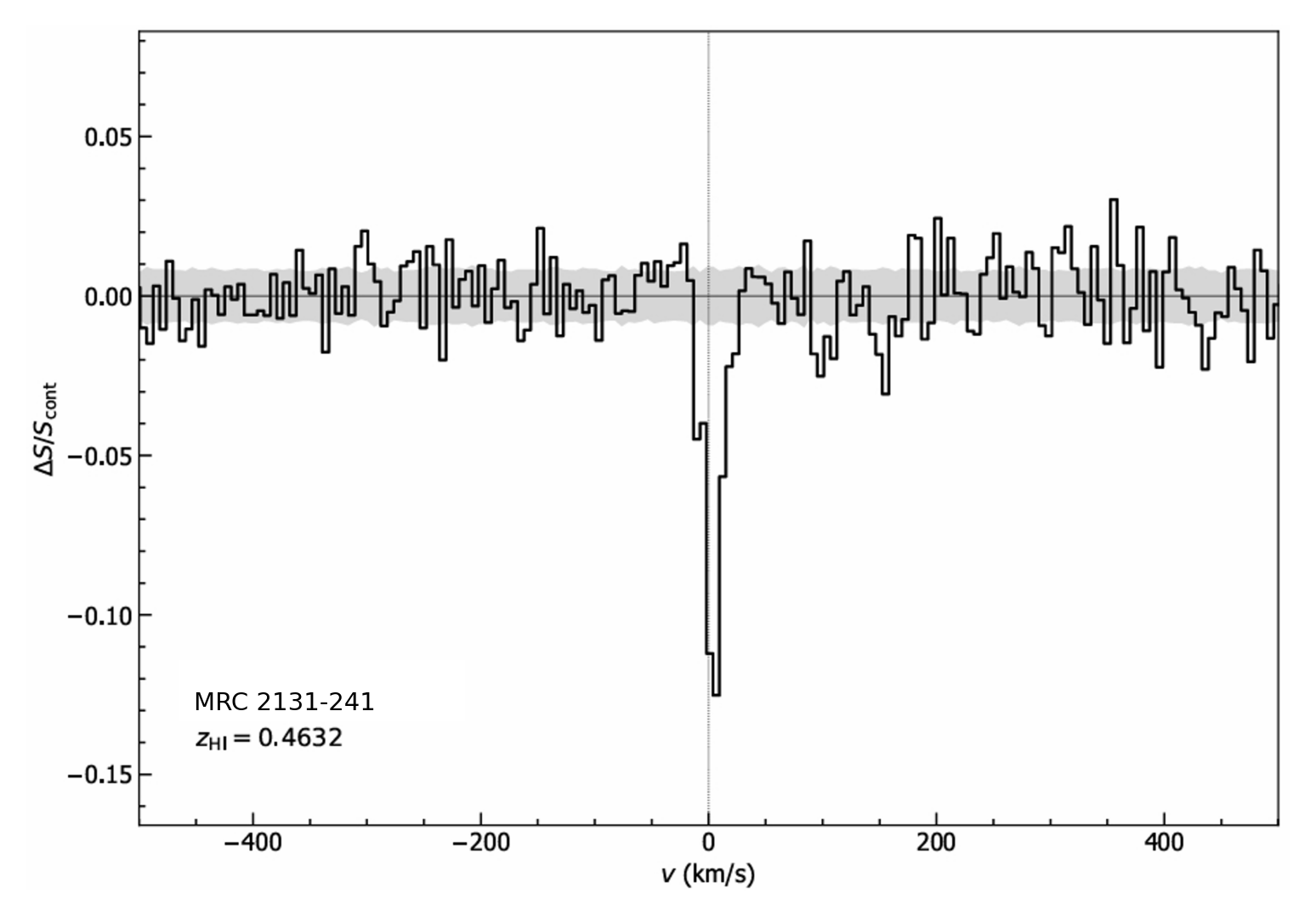}}
        \subfloat[FLASH ID SB66727 12a]{\includegraphics[width=0.5\textwidth]{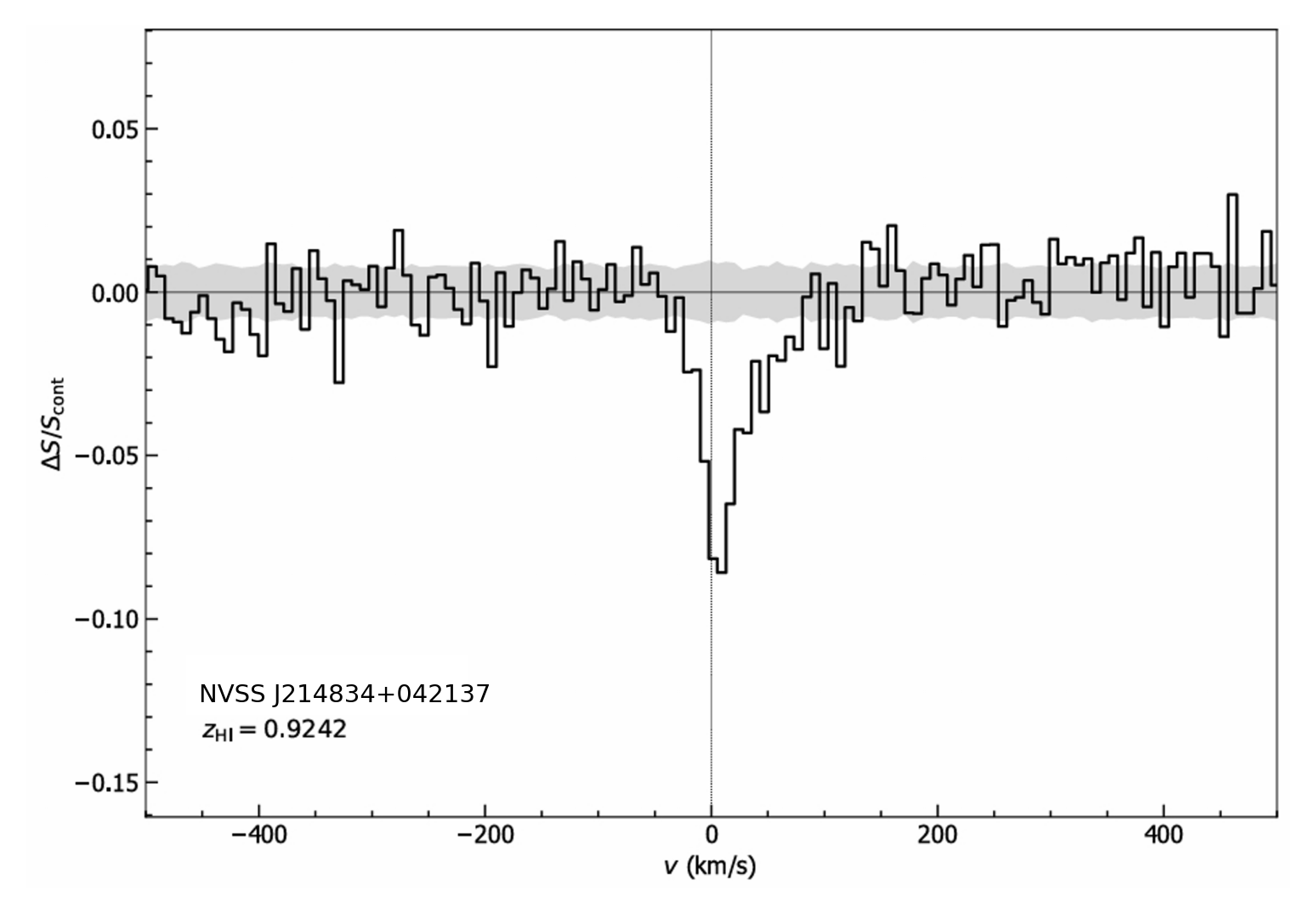}}
        \end{figure}

\end{document}